\documentclass[11pt,epsf]{article}
 \usepackage{amsmath}
 \usepackage{graphicx}
 \usepackage[merge,numbers,compress]{natbib}
 \usepackage[T1]{fontenc}
 \usepackage{booktabs}
 \usepackage{xcolor} 
 \usepackage{xspace}
 \usepackage{dcolumn}
 \usepackage{placeins}
 \usepackage{amssymb}
 \usepackage[colorlinks=true,citecolor=blue!50!black,linkcolor=black]{hyperref}
 \usepackage{caption}
 \usepackage
 [subrefformat=parens,position=top,skip=-15pt,margin=15pt,justification=justified,singlelinecheck=false]
 {subcaption}
 \usepackage{array,multirow}

\newcommand{\Pl}{\ell}

\newcommand{\pb}{{\ensuremath\unskip\,\text{pb}}\xspace}

\def\be{\begin{equation}}
\def\ee{\end{equation}}

\newcommand{\Pj}{\ensuremath{\text{j}}\xspace}
\newcommand{\Pp}{\ensuremath{\text{p}}\xspace}

\newcommand{\Pb}{\ensuremath{\text{b}}\xspace}
\newcommand{\Pq}{\ensuremath{{q}}\xspace}
\newcommand{\Pt}{\ensuremath{\text{t}}\xspace}
\newcommand{\Pu}{\ensuremath{\text{u}}\xspace}
\newcommand{\Pd}{\ensuremath{\text{d}}\xspace}
\newcommand{\Ps}{\ensuremath{\text{s}}\xspace}
\newcommand{\Pc}{\ensuremath{\text{c}}\xspace}
\newcommand{\Pg}{\ensuremath{\text{g}}\xspace}

\newcommand{\PW}{\ensuremath{\text{W}}\xspace}
\newcommand{\PZ}{\ensuremath{\text{Z}}\xspace}

\newcommand{\MWOS}{\ensuremath{M_\PW^\text{OS}}\xspace}
\newcommand{\MW}{\ensuremath{M_\PW}\xspace}
\newcommand{\MZOS}{\ensuremath{M_\PZ^\text{OS}}\xspace}
\newcommand{\MZ}{\ensuremath{M_\PZ}\xspace}

\newcommand{\GZOS}{\ensuremath{\Gamma_\PZ^\text{OS}}\xspace}

\newcommand{\GWOS}{\ensuremath{\Gamma_\PW^\text{OS}}\xspace}

\newcommand{\GeV}{\ensuremath{\,\text{GeV}}\xspace}
\newcommand{\TeV}{\ensuremath{\,\text{TeV}}\xspace}

\newcommand{\alphas}{\ensuremath{\alpha_\text{s}}\xspace}

\newcommand{\GF}{\ensuremath{G_\mu}}

\newcommand{\ptsub}[1]{\ensuremath{p_{\text{T},#1}}\xspace}

\newcommand{\MVOS}{\ensuremath{M_{\text{V}}^\text{OS}}\xspace}%
\newcommand{\GVOS}{\ensuremath{\Gamma_{\text{V}}^\text{OS}}\xspace}%

\newcommand{\newc}{\newcommand}
\newc{\bi}{\begin{itemize}}
\newc{\ei}{\end{itemize}}
\newc{\benu}{\begin{enumerate}}
\newc{\eenu}{\end{enumerate}}
\newc{\bc}{\begin{center}}
\newc{\ec}{\end{center}}
\newc{\bfig}{\begin{figure}}
\newc{\efig}{\end{figure}}
\newc{\qbar}{\bar{q}}
\newc{\go}{\tilde{g}}
\newc{\PB}{\textsc{Powheg-Box}}

\newcommand{\rT}{{\mathrm{T}}}
\newcolumntype{.}{D{.}{.}{-1}}
\newcolumntype{d}[1]{D{.}{.}{#1}}

\newcommand{\QCD}{\ensuremath{\text{QCD}}}

\newcommand{\mr}[1]{\ensuremath{\mathrm{#1}}}

\newcommand{\muR}{\ensuremath{\mu_{\mr{R}}}}
\newcommand{\muF}{\ensuremath{\mu_{\mr{F}}}}

\colorlet{tableoverheadcolor}{gray!37.5}
\colorlet{tableheadcolor}{gray!25}
\colorlet{tablerowcolor}{gray!12.5}

\newlength{\width}
\newlength{\height}

\def\draftdate{\relax}
\def\mda{\relax}
\def\mua{\relax}
\def\mla{\relax}
\def\draft{
\def\thtystars{******************************}
\def\sixtystars{\thtystars\thtystars}
\typeout{}
\typeout{\sixtystars**}
\typeout{* Draft mode!
         For final version remove \protect\draft\space in source file *}
\typeout{\sixtystars**}
\typeout{}
\def\draftdate{\today}
\def\mua{\marginpar[\boldmath\hfil$\uparrow$]%
                   {\boldmath$\uparrow$\hfil}\color{black}%
                    \typeout{marginpar: $\uparrow$}\ignorespaces}
\def\mda{\color{red}\marginpar[\boldmath\hfil$\downarrow$]%
                   {\boldmath$\downarrow$\hfil}%
                    \typeout{marginpar: $\downarrow$}\ignorespaces}
\def\mla{\marginpar[\boldmath\hfil$\rightarrow$]%
                   {\boldmath$\leftarrow $\hfil}%
                    \typeout{marginpar: $\leftrightarrow$}\ignorespaces}
\def\Mua{\marginpar[\boldmath\hfil$\Uparrow$]%
                   {\boldmath$\Uparrow$\hfil}\color{black}%
                    \typeout{marginpar: $\uparrow$}\ignorespaces}
\def\Mda{\color{red}\marginpar[\boldmath\hfil$\Downarrow$]%
                   {\boldmath$\Downarrow$\hfil}%
                    \typeout{marginpar: $\downarrow$}\ignorespaces}
\def\Mla{\marginpar[\boldmath\hfil\textcolor{red}{$\Rightarrow$}]%
                   {\boldmath\textcolor{red}{$\Leftarrow $}\hfil}%
                    \typeout{marginpar: $\leftrightarrow$}\ignorespaces}
\overfullrule 5pt
\oddsidemargin 15mm
\marginparwidth 29mm
}

\begin{document}

\title{\hfill ~\\[-30mm]
\phantom{h} \hfill\mbox{\small CAVENDISH--HEP--20/13, FR-PHENO-2020-015, TTK-20-34, P3H-20-057}
\\[1cm]
\vspace{13mm}   \textbf{NNLO QCD predictions for W+c-jet production \\ at the LHC}}

\date{}
\author{
Micha\l{} Czakon$^{1\,}$\footnote{E-mail:  \texttt{mczakon@physik.rwth-aachen.de}},
Alexander Mitov$^{2\,}$\footnote{E-mail:  \texttt{adm74@cam.ac.uk}},
Mathieu Pellen$^{2\, , 3\,}$\footnote{E-mail:  \texttt{mathieu.pellen@physik.uni-freiburg.de}},
Rene Poncelet$^{2\,}$\footnote{E-mail:  \texttt{poncelet@hep.phy.cam.ac.uk}}
\\[9mm]
{\small\it $^1$ Institut f{\"u}r Theoretische Teilchenphysik und Kosmologie, RWTH Aachen University,} \\ %
{\small\it RWTH Aachen University, D-52056 Aachen, Germany}\\[3mm]
{\small\it $^2$ Cavendish Laboratory, University of Cambridge,} \\ %
{\small\it J.J. Thomson Avenue, Cambridge CB3 0HE, United Kingdom}\\[3mm]
{\small\it $^3$ Albert-Ludwigs-Universit\"at Freiburg, Physikalisches Institut,} \\ %
{\small\it Hermann-Herder-Stra\ss e 3, D-79104 Freiburg, Germany}\\[3mm]
        }
\maketitle

\begin{abstract}
\noindent

We study the production of a W boson in association with a c-jet at the LHC. We calculate, for the first time, the complete set of NNLO QCD corrections to the dominant CKM-diagonal contribution to this process.
Both signatures, $\Pp\Pp\to \mu^+\nu_\mu\Pj_{\rm c}$ and $\Pp\Pp\to \mu^-\bar\nu_\mu\Pj_{\rm c}$ are considered.
We present predictions for fiducial cross sections and differential distributions for each one of the two signatures as well as for their ratio.
The theoretical predictions are compared with ATLAS measurements at $7\TeV$.
The results of this work are essential for the precision description of associated heavy flavor production at hadron colliders and for the determination of the strange-quark content of the proton from LHC data in NNLO QCD.

\end{abstract}
\thispagestyle{empty}
\vfill

\newpage

\section{Introduction}

The Large Hadron Collider (LHC) has triggered an unprecedented number of high-precision studies of the electroweak (EW) and strongly interacting sectors of the Standard Model (SM) as well as searches for physics beyond the SM.
Despite all this progress in precision physics, the heavy-flavor (bottom and charm) sector of the SM has so far benefited relatively little in terms of high-precision theoretical work.
While the (experimentally inaccessible) total inclusive cross section for $b\bar b$ and $c\bar c$ production can be inferred in NNLO QCD from existing $\Pt\bar \Pt$ calculations \cite{Baernreuther:2012ws,Czakon:2012zr,Czakon:2012pz,Czakon:2013goa} no differential observable related to $\Pb$ or $\Pc$ production was known at NNLO in QCD until very recently.
This situation can be contrasted with light-flavor production whose description in NNLO QCD has been established years ago by several groups in W+jet processes \cite{Boughezal:2015dva,Boughezal:2016dtm,Gehrmann-DeRidder:2019avi} and inclusive jet production \cite{Currie:2016bfm,Currie:2017eqf,Currie:2018xkj,Gehrmann-DeRidder:2019ibf,Czakon:2019tmo}, as well with top-quark production which has also been known in NNLO QCD \cite{Czakon:2014xsa,Czakon:2015owf,Czakon:2016ckf,Czakon:2016dgf,Gao:2017goi,Behring:2019iiv,Catani:2019hip,Czakon:2020qbd,Brucherseifer:2014ama,Berger:2016oht,Berger:2017zof} for some time.

The first step towards computing heavy flavor production in NNLO QCD was achieved only very recently in ref.~\cite{Gauld:2020deh} where the NNLO QCD corrections for $\PZ+\Pb$ production at the LHC were computed
\footnote{While the present study was being prepared for publication a first NNLO QCD calculation of differential $\Pb\bar \Pb$ production at quark level was presented in ref.~\cite{Catani:2020kkl}.}.
With the help of an independent computational formalism in this work we calculate for the first time the NNLO QCD corrections for a LHC process involving charm quark, specifically, the associated production of a charm jet and a $\PW$ boson. 

The theoretical progress in $V$+jet processes has a long history. For the case of $\PW$ production with inclusive jets the next-to-leading-order (NLO) QCD corrections have been known for some time \cite{Giele:1993dj,Arnold:1988dp,Arnold:1989ub,Campbell:2002tg,Campbell:2008hh,Caola:2011pz}.
Pure EW corrections have been studied extensively \cite{Kuhn:2007qc,Kuhn:2007cv,Hollik:2007sq,Denner:2009gj}.
The combined EW and QCD corrections \cite{Kallweit:2014xda,Kallweit:2015dum,Biedermann:2017yoi} are also available.
The above state of the art predictions have then been combined in ref.~\cite{Lindert:2017olm} with the known NNLO QCD corrections for the purpose of a study on SM-backgrounds to Dark-Matter searches.
Prior studies of W+c-jet production in NLO QCD are available for both the Tevatron \cite{Giele:1995kr} and LHC \cite{Stirling:2012vh}.
Measurements of this process have been performed by both the ATLAS \cite{Aad:2014xca} and CMS \cite{CMS:2018muk,Sirunyan:2018hde,CMS:2019rlx} collaborations.

One of the main goals of the present calculation is to offer a high-precision access to the strange quark parton distribution function (PDF) which is one of the least constrained proton PDFs.
$\PW+\Pc$ production at the LHC is a powerful probe for the strange PDF \cite{Baur:1993zd} and its asymmetry \cite{Catani:2004nc} because at tree-level and for the dominant CKM matrix element it features a strange (anti-)quark and a gluon in its initial state.
References~\cite{Lai:2007dq,Stirling:2012vh,Yalkun:2019gah,Faura:2020oom} have studied charm production in the context of strange PDF determination in NLO QCD.

Another main goal of this study is to provide state-of-the-art QCD predictions for the process $\PW$+c-jet. More precisely, predictions are provided for the two signatures $\Pp\Pp\to \mu^+\nu_\mu\Pj_{\rm c}$ and $\Pp\Pp\to \mu^-\bar\nu_\mu\Pj_{\rm c}$ in terms of cross sections, differential distributions, and ratios.
Our NNLO QCD predictions include PDF uncertainties and estimates of missing higher-order terms.
These new predictions are also compared to the experimental results obtained by the ATLAS collaboration \cite{Aad:2014xca} for the fiducial cross section and the differential distributions in the rapidity of the (anti-)muon.
A number of other differential distributions for which there is no data are also presented for the purpose of assessing the quality of the theoretical description of this process.

This article is organized as follows. In sec.~\ref{sec:details} we present the definition of the processes and the details of the computational setup.
Section~\ref{sec:numerical-results} is devoted to the presentation and discussion of our numerical results and their comparison with existing experimental data.
Section~\ref{sec:conclusions} contains a summary of our main findings together with some concluding remarks.

\section{Details of the calculations}\label{sec:details}

\subsection{Definition of the process}

The process under current investigation is the off-shell production of a W boson in association with a c-jet at the LHC in its proton-proton collision mode.
The two signatures relevant for this processes are
\begin{eqnarray}
&& \Pp\Pp \to \mu^+  \nu_{\mu} \Pj_{\rm c} + X\,,\label{eq:w+j}\\
&& \Pp\Pp \to \mu^-  \bar\nu_{\mu} \Pj_{\rm c} + X .\label{eq:w-j}
\end{eqnarray}

In the following we sometimes refer to the above processes as $\PW^\pm\Pj_{\rm c}$ but the full off-shell processes as in eqs.~\eqref{eq:w+j} and \eqref{eq:w-j} are always meant.
At LO, assuming a diagonal CKM matrix, each one of the hadronic processes in eqs.~\eqref{eq:w+j} and \eqref{eq:w-j} involves a single partonic reaction, namely $\bar\Ps\Pg \to \mu^+ \nu_{\mu} \bar\Pc$ and $\Ps\Pg \to \mu^- \bar\nu_{\mu} \Pc$, respectively.
The partonic channels $\bar\Pd\Pg \to \mu^+ \nu_{\mu} \bar\Pc$ and $\Pd\Pg \to \mu^- \bar\nu_{\mu} \Pc$ contribute, too, once the off-diagonal CKM matrix elements are taken into account
\footnote{The CKM matrix element $V_{\Pc\Pb}$ is always neglected.}.

The above processes eqs.~\eqref{eq:w+j} and \eqref{eq:w-j} are defined at order $\mathcal{O} \left(\alphas \alpha^2 \right)$ in the strong and EW couplings.
In the following, our best predictions feature NNLO QCD corrections to the CKM-diagonal channel $\bar\Ps\Pg \to \mu^+ \nu_{\mu} \bar\Pc$/$\Ps\Pg \to \mu^- \bar\nu_{\mu} \Pc$
and include the non-diagonal CKM channel $\bar\Pd\Pg \to \mu^+ \nu_{\mu} \bar\Pc$/$\Pd\Pg \to \mu^- \bar\nu_{\mu} \Pc$ at LO in QCD.
This means that our best predictions include the effects of $V_{\Pc\Ps}\neq0$ through NNLO QCD while the ones of $V_{\Pc\Pd}\neq0$ only at LO in QCD.

The NLO QCD corrections are of order $\mathcal{O} \left(\alphas^2 \alpha^2 \right)$.
The real corrections consist of all partonic processes that have in the final state a (decaying) W boson in association with two partons, at least one of which is a charm quark or anti-quark.
These partonic reactions are made of either four external quarks or two quarks and two gluons.
This implies that beyond LO, the initial state consists not only of a gluon and a strange quark but could be one of several other parton pairs. 
The NNLO corrections are of order $\mathcal{O} \left(\alphas^3 \alpha^2 \right)$ and feature the opening of yet new partonic channels in the double-real radiation part of the computation.
In order to quantify the numerical importance of the various initial-state channels, we have split the calculations in terms of 8 possible channels. 
These are listed in Table~\ref{tab:contributions} together with the perturbative order at which they first contribute.
Note that these categories are infrared (IR) finite since they are separated according to their initial-state flux.
It is worth noticing that already at NLO almost all partonic reactions contribute, apart from the $\Pg\Pq$ channel ($\Pq$ stands for any quark or anti-quark which is not $s$ or $\bar s$) and 
$\Ps\Pg$ ($\bar\Ps\Pg$) for $\Pp\Pp \to \PW^+\Pj_{\rm c}$ ($\Pp\Pp \to \PW^-\Pj_{\rm c}$).
These partonic channels only contribute at NNLO in the CKM-diagonal case.

\begin{table}
\centering
\begin{tabular}{cccc}
\multicolumn{4}{c}{\raisebox{0.2cm}{$\Pp\Pp \to \PW^+\Pj_{\rm c}$}} \\
\toprule
Contribution    & LO    & NLO    & NNLO \\
\midrule
    $\bar\Ps\Pg$    &  {\bf \checkmark}    &  {\bf \checkmark}    &  {\bf \checkmark} \\
    $\Ps\Pg$    &  {X}    &  {X}    &  {\bf \checkmark} \\
    $\Ps\bar\Ps$    & {X}    & {\bf \checkmark}    & {\bf \checkmark} \\
    $\bar\Ps\bar\Ps$    & {X}    & {\bf \checkmark}    & {\bf \checkmark} \\
    $\bar\Ps\Pq$    & {X}    & {\bf \checkmark}    & {\bf \checkmark} \\
    $\Pq\Pq'$    & {X}    & {\bf \checkmark}    & {\bf \checkmark} \\
    $\Pg\Pq$    & {X}    & {X}    & {\bf \checkmark} \\
    $\Pg\Pg$    & {X}    & {\bf \checkmark}    & {\bf \checkmark} \\
\bottomrule
\end{tabular}
\centering
\hspace{1cm}
\begin{tabular}{cccc}
\multicolumn{4}{c}{\raisebox{0.2cm}{$\Pp\Pp \to \PW^-\Pj_{\rm c}$}} \\
\toprule
Contribution    & LO    & NLO    & NNLO \\
\midrule
    $\bar\Ps\Pg$    &  {X}    &  {X}    &  {\bf \checkmark} \\
    $\Ps\Pg$    &  {\bf \checkmark}    &  {\bf \checkmark}    &  {\bf \checkmark} \\
    $\Ps\bar\Ps$    & {X}    & {\bf \checkmark}    & {\bf \checkmark} \\
    $\Ps\Ps$    & {X}    & {\bf \checkmark}    & {\bf \checkmark} \\
    $\Ps\Pq$    & {X}    & {\bf \checkmark}    & {\bf \checkmark} \\
    $\Pq\Pq'$    & {X}    & {\bf \checkmark}    & {\bf \checkmark} \\
    $\Pg\Pq$    & {X}    & {X}    & {\bf \checkmark} \\
    $\Pg\Pg$    & {X}    & {\bf \checkmark}    & {\bf \checkmark} \\
\bottomrule
\end{tabular}
\caption{List of initial-state contributions for $\Pp\Pp \to \PW^+\Pj_{\rm c}$ and $\Pp\Pp \to \PW^-\Pj_{\rm c}$ that are present at LO, NLO, and NNLO QCD.
Any (anti-)quark that is not a (anti-)strange quark is denoted by $\Pq$ or $\Pq'$.
Here $V_{\Pc\Pd}=0$ is assumed.}
\label{tab:contributions}
\end{table}

\subsection{Numerical inputs}

The predictions are obtained for proton-proton collisions at the LHC running at a centre-of-mass energy of $\sqrt{s} = 7 \TeV$.
The 5-flavor scheme is used throughout the computation \emph{i.e.}\
the bottom quarks are considered massless.
Bottom quarks are part of the light jets and thus do not carry flavor as opposed to the charm quarks (see below).
In this calculation we account for the fact that the CKM matrix is different from the unit one, thus including quark mixing effects.
The values of the entries of the CKM matrix used for the numerical simulations are taken from the PDG \cite{Tanabashi:2018oca} using the global fit numbers
\begin{equation}
V_{\Pc\Ps} = 0.97359 \quad \text{and} \quad V_{\Pc\Pd} = 0.22438 .
\end{equation}

The PDF sets used for the LO, NLO, and NNLO computations are the NNPDF31 sets with $\alphas=0.118$ \cite{Ball:2017nwa} matching the corresponding orders.
The strong coupling $\alpha_s$ is also extracted from there using LHAPDF6~\cite{Buckley:2014ana}.
In order to present PDF variation at NNLO QCD, we have used specialised minimal PDF sets \cite{Carrazza:2016htc}.
This means that instead of computing the NNLO results for the 100 replicas of the original set, we have used only about ten of these reduced set for each of the processes eqs.~\eqref{eq:w+j} and \eqref{eq:w-j}.
This leads to a significant gain in computing efficiency without loss of information related to PDF uncertainty.

The electromagnetic coupling is obtained in the $G_\mu$ scheme \cite{Denner:2000bj} using the Fermi constant
\begin{equation}
  \alpha = \frac{\sqrt{2}}{\pi} G_\mu \MW^2 \left( 1 - \frac{\MW^2}{\MZ^2} \right)  \qquad \text{with}  \qquad   {\GF = 1.16638\times 10^{-5}\GeV^{-2}}.
\end{equation}

The values of the masses and widths used for the numerical simulations read \cite{Tanabashi:2018oca}
\begin{alignat}{2}
\label{eqn:ParticleMassesAndWidths}
                \MZOS &=  91.1876\GeV,      & \quad \quad \quad \GZOS &= 2.4952\GeV,  \nonumber \\
                \MWOS &=  80.379\GeV,       & \GWOS &= 2.085\GeV.
\end{alignat}
The pole masses and widths used in the calculation are obtained from the measured on-shell (OS) values \cite{Bardin:1988xt} for the massive gauge bosons using
\begin{equation}
        M_V = \frac{\MVOS}{\sqrt{1+(\GVOS/\MVOS)^2}}\,,\qquad  
\Gamma_V = \frac{\GVOS}{\sqrt{1+(\GVOS/\MVOS)^2}}.
\end{equation}

In all computations, the intermediate W-boson resonances are treated in the complex-mass scheme~\cite{Denner:1999gp,Denner:2005fg,Denner:2006ic} to ensure the gauge invariance of all amplitudes.

Finally, the common central renormalization ($\muR$) and factorization ($\muF$) scale used for the present computation is
\begin{equation}
\label{scale}
 \mu = \frac12 \left( E_{\text{T},W} + \ptsub{\Pj_\text{c}} \right),
\end{equation}
where $E_{\rm T, \PW} = \sqrt{\MW^2 + \left( \vec{p}_{\rm T,\Pl} + \vec{p}_{\rm T, \nu} \right)^2}$.
Closely related scale definitions have been used in the past for V+jet processes (see \emph{e.g.}\ refs.~\cite{Bertone:2017djs,Gehrmann-DeRidder:2019avi,Gauld:2020deh} and references therein).
The scale uncertainty of fiducial cross sections and differential distributions is obtained by taking the envelope of the 7-point variations of the renormalization and factorization scale \emph{i.e.}\ 
$\{(\tfrac{1}{2}\muR,\tfrac{1}{2}\muF)$, $(\tfrac{1}{2}\muR,\muF)$, $(\muR,\tfrac{1}{2}\muF)$, $(\muR,\muF)$, $(\muR,2\muF)$, $(2\muR,\muF)$, $(2\muR,2\muF)\}$.

\subsection{Event selections and flavored jet algorithm}
\label{sec:cuts}
 
The event selections used in the present computation follow closely the experimental ones for the W+c-jet ATLAS analysis of ref.~\cite{Aad:2014xca}.
We reproduce them below for completeness.
The final state considered is a c-jet in association with a charged (anti-)muon as well as missing transverse energy.
The requirement on the charged lepton reads
\begin{align}
 \ptsub{\ell} >  20\GeV, \qquad |\eta_\ell| < 2.5.
\end{align}
In addition, each event is required to fulfil
\begin{align}
 p_{\mathrm{T},\text{miss}} >  25\GeV \qquad \text{and} \qquad m_{\rm T}^{\rm W} > 40\GeV.
\end{align}
In our calculation $p_{\mathrm{T},\text{miss}}$ is defined as the transverse momentum of the neutrino.
The W-boson transverse mass reads
\begin{equation}
m_{\rm T}^{\rm W} = \sqrt{2 \ptsub{\ell} p_{\mathrm{T},\text{miss}} \left(1 - \cos \Delta \phi \right)}\,,
\label{eq:mTW} 
\end{equation}
with $\Delta \phi = \min \left( |\phi_\ell - \phi_\nu|, 2\pi - |\phi_\ell - \phi_\nu| \right)$ being the azimuthal-angle separation of the lepton and neutrino momenta.

The c-jets are obtained by applying the flavored $k_\rT$ algorithm \cite{Banfi:2006hf} as implemented in ref.~\cite{Gauld:2019yng} with a jet-resolution parameter of $R=0.4$.
As opposed to the standard (anti-)$k_\rT$ algorithm, the distance between pseudo-jets $i$ and $j$ ($d_{ij}$) is dependent on the flavor of the considered partons

\begin{equation}
  d_{ij} = \frac{\Delta {y_{ij}^2}+\Delta {\phi_{ij}^2}}{R^2}
    \begin{cases}
      max\left( k_{\rT i}, k_{\rT j} \right)^2 & \text{if softer of $i$,$j$ is flavored} \\
      min\left( k_{\rT i}, k_{\rT j} \right)^2 & \text{if softer of $i$,$j$ is unflavored}
    \end{cases}
\end{equation}
which corresponds to the case $\alpha=2$ in the original formulation \cite{Banfi:2006hf}.
The distance to the beam is also flavor dependent and is defined as
\begin{equation}
  d_{i\beta} = 
    \begin{cases}
      max\left( k_{\rT i}, k_{\rT \beta} \left(y_i\right) \right)^2 & \text{if $i$ is flavored} \\
      min\left( k_{\rT i}, k_{\rT \beta} \left(y_i\right) \right)^2 & \text{if $i$ is unflavored} 
    \end{cases}
\end{equation}
with $\beta = B, \bar{B}$. The beam transverse momentum is rapidity-dependent 
\begin{alignat}{2}
  k_{\rT B} \left(y\right) &=& \sum_i k_{\rT i}\left( \Theta \left(y_i - y \right) + \Theta \left(y - y_i \right) \exp{\left(y_i-y\right)} \right), \\
  k_{\rT \bar{B}} \left(y\right) &=& \sum_i k_{\rT i}\left( \Theta \left(y - y_i \right) + \Theta \left(y_i - y \right) \exp{\left(y-y_i\right)} \right),
\end{alignat}
with the index $i$ running over all pseudo-jets and $\Theta \left(0\right) = {\color{red} 1/2}$
\footnote{As opposed to ref.~\cite{Gauld:2019yng}, the beam measure does not include the W boson.}
.

At each step of the algorithm the three distances $d_{ij}$, $d_{i B}$, and $d_{i \bar{B}}$ have to be computed.
Following ref.~\cite{Gauld:2019yng}, jets with an even number of charm quarks are declared unflavored (for example $\Pc\bar\Pc$ or $\Pc\Pc$ pairs) while
jets with an odd number of charm quarks are declared flavored (for example $\Pc\Pc\bar\Pc$ or $\bar\Pc\Pc\bar\Pc$ triplets).

For an event to be accepted, one and only one flavored c-jet should fulfil the following criteria:
\begin{align}
\label{eq:jet}
 \ptsub{\Pj_c} >  25\GeV, \qquad |\eta_{\Pj_c}| < 2.5.
\end{align}
If two (or more) c-jets fulfil eq.~\eqref{eq:jet}, the event is rejected.

\subsection{Description of the calculation and its validation}

The computation has been carried out within {\sc Stripper}, a c++ implementation of the four-dimensional formulation of the sector-improved residue subtraction scheme \cite{Czakon:2010td,Czakon:2011ve,Czakon:2014oma}.
This framework has already been applied in NNLO QCD to the production of top-quark pairs \cite{Czakon:2014xsa,Czakon:2015owf,Czakon:2016ckf,Behring:2019iiv,Czakon:2020qbd}, inclusive jets \cite{Czakon:2019tmo} and three photons \cite{Chawdhry:2019bji}.
Details about the four-dimensional implementation of the subtraction scheme can be found in \cite{Czakon:2014oma,Czakon:2019tmo}.
While the full integration and subtraction of IR divergences is handled by {\sc Stripper}, it relies on external tools for tree-level, one-loop, and two-loop matrix elements.
In the present computation, the {\sc AvH} library \cite{Bury:2015dla} has been used for the tree-level part.
The one-loop matrix elements used in the real-virtual part of the computation have been taken from the library {\sc OpenLoops 2} \cite{Buccioni:2019sur}.
Finally, the one-loop and two-loop amplitudes for the $\Pp\Pp\to\PW+\Pj$ process have been obtained from ref.~\cite{Gehrmann:2011ab}. The library {\sc Ginac} \cite{Bauer:2000cp,Vollinga:2004sn}
has been used for the numerical evaluation of the harmonic polylogarithms (HPLs, \cite{Remiddi:1999ew}) and two-dimensional harmonic polylogarithms (2dHPLs, \cite{Gehrmann:2001ck}).

During the course of this computation, many cross checks have been performed.
We list the most important ones here.
The full $\PW+\Pj$ computation at NNLO QCD has been validated against the results of ref.~\cite{Gehrmann-DeRidder:2019avi} at the level of the fiducial cross section and differential distributions.
The flavored-jet algorithm has been implemented in two independent ways following refs.~\cite{Banfi:2006hf,Gauld:2019yng} which have been cross-checked for large flavored jets multiplicities.
The implementation of the (polarised) amplitude from ref.~\cite{Gehrmann:2011ab} has been checked against {\sc Recola} \cite{Actis:2016mpe,Actis:2012qn} at tree and one-loop level and the evaluation of the harmonic polylogarithms using {\sc Ginac} has been checked against the libraries {\sc hplog} \cite{Gehrmann:2001pz} and {\sc tdhpl} \cite{Gehrmann:2001jv}.
Finally, the LO and NLO QCD predictions for $\PW+\Pj_{\rm c}$ have been successfully compared against an independent Monte Carlo program, 
{\sc MoCaNLO}+{\sc Recola} which has already been used in several V+jets computations \cite{Biedermann:2016yds,Biedermann:2017bss,Ballestrero:2018anz,Chiesa:2019ulk,Denner:2019tmn,Denner:2019zfp,Pellen:2019ywl,Brauer:2020kfv,Denner:2020zit}.
This last check also validates the implementation of the event selection as well as the correct use of the flavored jet algorithm up to NLO QCD accuracy.

\section{Numerical results}\label{sec:numerical-results}

\subsection{Cross sections}\label{sec:cross-section}

In this section, cross sections at LO, NLO, and NNLO QCD accuracy are presented for $\Pp\Pp \to \PW^+\Pj_{\rm c}$ and $\Pp\Pp \to \PW^-\Pj_{\rm c}$ at $7\TeV$ at the LHC.
The ratio of the two cross sections is also provided.
The theoretical predictions are compared with measurements of the ATLAS collaboration \cite{Aad:2014xca}.

\begin{table}
  \begin{center}
    \begin{tabular}{c|c|c|c}
    \multicolumn{4}{c}{\raisebox{0.2cm}{ $V_{\Pc\Pd}\neq0$}} \\
    \toprule
    Order & $\sigma_{\PW^+\Pj_{\rm c}}$ [$\pb$] & $\sigma_{\PW^-\Pj_{\rm c}}$ [$\pb$] & $R_{\PW^\pm\Pj_{\rm c}} = \sigma_{\PW^+\Pj_{\rm c}}/\sigma_{\PW^-\Pj_{\rm c}}$\\
    \midrule
    LO    & $12.0725(4)^{+11.6\%}_{-12.9\%}$      & $14.2624(5)^{+11.6\%}_{-10.9\%}$ & $0.84646(4)^{+1.48\%}_{-2.22\%}$ \\
    \midrule
    NLO   & $35.164(9)^{+8.0\%}_{-7.0\%}$        & $37.096(9)^{+7.5\%}_{-6.7\%}$     & $0.9479(3)^{+0.49\%}_{-0.36\%}$ \\
    \midrule
    NNLO  & $38.6(1)^{+2.2\% \; +3.8\% ({\rm PDF})}_{-3.2\% \; -3.8\% ({\rm PDF})}$ & $39.3(1)^{+1.8\% \; +3.9\% ({\rm PDF})}_{-2.9\% \; -3.9\% ({\rm PDF})}$ 
    & $0.983(5)^{+0.45\%\; +2.7\% ({\rm PDF})}_{-0.37\%\; -2.7\% ({\rm PDF})}$ \\
    \bottomrule
    \end{tabular}
  \end{center}
  \caption{\label{tab:wcCKM}
    Fiducial cross sections for $\Pp\Pp \to \PW^+\Pj_{\rm c}$, $\Pp\Pp \to \PW^-\Pj_{\rm c}$, and their ratios at the LHC at $\sqrt{s}=7\TeV$ at LO, NLO, and NNLO \QCD.
    The digit in parenthesis indicates the Monte Carlo statistical error while the sub- and super-script in per cent indicate the scale variation.
    In addition, the PDF variation is provided for the NNLO QCD predictions (as indicated explicitly).
    The contribution due to $V_{\Pc\Pd}\neq0$ is included at Born level.
    The NNPDF3.1 sets with $\alphas=0.118$ are used at orders matching the perturbative ones.}
\end{table}

Our best predictions for the fiducial cross section are presented in Table~\ref{tab:wcCKM}.
As explained previously the NLO and NNLO QCD corrections are computed only for the CKM-diagonal channel $\bar\Ps\Pg \to \mu^+ \nu_{\mu} \bar\Pc$/$\Ps\Pg \to \mu^- \bar\nu_{\mu} \Pc$
while the non-diagonal CKM channel $\bar\Pd\Pg \to \mu^+ \nu_{\mu} \bar\Pc$/$\Pd\Pg \to \mu^- \bar\nu_{\mu} \Pc$ is included at LO in QCD.

The first interesting point is that the NLO QCD corrections are extremely large, about $200\%$.
These gigantic corrections are largely driven by the different PDF sets used at LO and NLO where both the gluon and (anti-)strange PDF vastly differ.
If one uses the same NNLO PDF at each perturbative order the corrections are still rather large but reduce greatly to about $(40-50)\%$.
On the other hand, the NNLO QCD corrections are smaller than the NLO ones and do not exceed $10\%$ ($+8.9\%$ for the plus signature and $+5.6\%$ for the minus one)
\footnote{We note that the NLO and NNLO PDF sets only differ by few per cent.}.
This observation is in line with previous $\PW+\Pj$ computations \cite{Boughezal:2015dva,Boughezal:2016dtm,Gehrmann-DeRidder:2019avi} with no flavor tagging.
This pattern of higher-order corrections can be explained by new topologies appearing at NLO such as di-jet topologies with soft-collinear W radiations.
Such configurations lead to so-called giant K-factors \cite{Rubin:2010xp}. At NNLO QCD no such new topologies appear explaining the relatively small NNLO corrections.

From Table~\ref{tab:wcCKM} we also conclude that the scale variation of the fiducial cross section is strongly affected by the inclusion of higher-order corrections. While at LO it is about $\pm(10-15)\%$ the scale variation decreases to a few per cent at NNLO QCD.

In the third column of Table~\ref{tab:wcCKM}, the ratio of the cross sections for the two signatures is provided. It is defined as 
\begin{equation}
 R_{\PW^\pm\Pj_{\rm c}} = \frac{\sigma_{\PW^+\Pj_{\rm c}}}{\sigma_{\PW^-\Pj_{\rm c}}}\,.
\end{equation}
In computing the scale variation of $R_{\PW^\pm\Pj_{\rm c}}$ the scale uncertainties between the two signatures are taken correlated.
This ratio tends to get closer to 1 when including higher order corrections and changes by about $+15\%$ when going from LO to NNLO accuracy.
Since at LO not only the $\bar\Ps\Pg$ ($\Ps\Pg$) but also the $\bar\Pd\Pg$ ($\Pd\Pg$) initial state contributions are included, 
the ratio $R_{\PW^\pm\Pj_{\rm c}}$ behaves schematically as $\left(|V_{\Pc\Ps}|\bar\Ps +|V_{\Pc\Pd}|\bar \Pd \right)/\left(|V_{\Pc\Ps}|\Ps +|V_{\Pc\Pd}|\Pd \right)$.
The inclusion of the non-diagonal CKM channel $\bar\Pd\Pg \to \mu^+ \nu_{\mu} \bar\Pc$/$\Pd\Pg \to \mu^- \bar\nu_{\mu} \Pc$ significantly lowers the ratio $R_{\PW^\pm\Pj_{\rm c}}$ due to the large asymmetry between down and anti-down quarks in the proton.
The impact of this channel on the ratio is diluted by the inclusion of higher-order corrections which explains why the inclusion of higher-order corrections brings the ratio closer to 1.

Alternatively, we present the ratio computed with {\it uncorrelated} scale uncertainties, that are obtained from a 31-point restricted scale variation. The results read
\begin{equation}
\label{eq:xsecUnc}
 R_{\PW^\pm\Pj_{\rm c}}^{\rm LO, unc.}   = 0.84646(4)^{+25.4\%}_{-22.0\%} , \quad 
 R_{\PW^\pm\Pj_{\rm c}}^{\rm NLO, unc.}  = 0.9479(3)^{+9.8\%}_{-8.6\%} , \quad 
 R_{\PW^\pm\Pj_{\rm c}}^{\rm NNLO, unc.} = 0.983(5)^{+3.5\%}_{-3.7\%} .
\end{equation}
We have checked that beyond LO, the 31-point scale variation produces variations very similar to the ones based on naive scale variation propagation.
As has already been observed in many processes, the correlated scale variation of ratios of observables is often significantly smaller than the uncorrelated one.
Such restricted variation likely underestimates the theoretical uncertainty in this ratio, at least at lower perturbative orders.

The PDF uncertainty of the NNLO QCD predictions is calculated according to ref.~\cite{Carrazza:2016htc}.
We observe that at this order the effect from missing-higher orders estimated via scale variation is smaller than the PDF uncertainty, in some cases by up to a factor of two.
The PDF uncertainty of the ratio has been computed according to eqs.~(35) and (40) of ref.~\cite{Carrazza:2016htc}.
The idea is to first map the results obtained with the reduced sets (which are different for each process) to the original replicas and then take their correlated ratio.
This procedure leads to a $\pm 2.7\%$ PDF uncertainty on $R_{\PW^\pm\Pj_{\rm c}}$, to be compared to the $\pm3.8\%$ or $\pm3.9\%$ PDF uncertainty on the two fiducial cross-sections.
Interestingly, the reduction of PDF uncertainty between the ratio and absolute cross sections is only about a factor of 2 compared to the reduction by a factor of 4-8 observed for the scale variation.

\begin{figure}
        \setlength{\parskip}{-10pt}
        \begin{subfigure}{0.98\textwidth}
                \subcaption{}
                \hspace{-10pt}
                 \includegraphics[width=\textwidth,page=1]{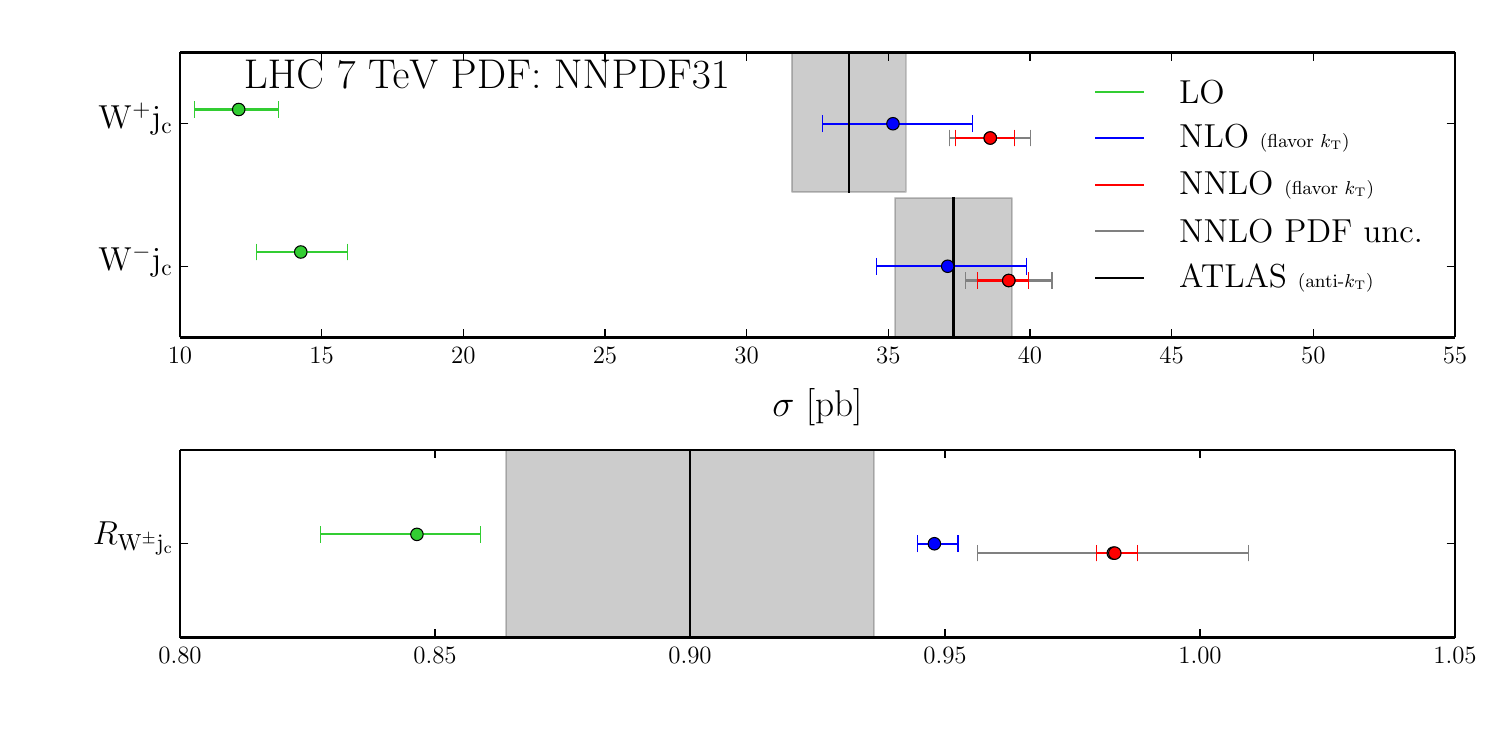}
        \end{subfigure}
        \vspace*{-3ex}
        \caption{\label{fig:xsection}%
                Cross sections for $\Pp\Pp \to \PW^+\Pj_{\rm c}$, $\Pp\Pp \to \PW^-\Pj_{\rm c}$, and the ratio $R_{\PW^\pm\Pj_{\rm c}}$ at the LHC with $\sqrt{s}=7\TeV$.
                The theoretical predictions are at LO (green), NLO (blue), and NNLO (red) in QCD and are compared to the ATLAS data (black) of ref.~\cite{Aad:2014xca}.
                The green/blue/red error bars represent scale variation. The PDF variation (grey error bars) is shown only at NNLO.
                The uncertainty of data represents the combined statistic and systematic error.
                The contribution due to $V_{\Pc\Pd}\neq0$ is included at Born level.
                The theory predictions are obtained using the flavor-$k_{\rm T}$ algorithm and
                the ATLAS data has been measured using anti-$k_{\rm T}$.
                }
\end{figure}

The fiducial cross sections measured by the ATLAS collaboration \cite{Aad:2014xca} are reproduced here for completeness and can be compared to the results of Table~\ref{tab:wcCKM}
\begin{eqnarray}
&& \sigma^{\rm ATLAS}_{\PW^+\Pj_{\rm c}} = 33.6 \pm 0.9 \;(\text{stat}) \pm 1.8 \;(\text{syst}) \pb\,,\\
&&\nonumber \\
&& \sigma^{\rm ATLAS}_{\PW^-\Pj_{\rm c}} = 37.3 \pm 0.8 \;(\text{stat}) \pm 1.9 \;(\text{syst}) \pb \,.
\end{eqnarray}
In addition, the measured ratio of these two cross sections is also provided
\begin{equation}
 R^{\rm ATLAS}_{\PW^\pm\Pj_{\rm c}} = 0.90 \pm 0.03 \;(\text{stat}) \pm 0.02 \;(\text{syst}) \,.
\end{equation}
The theory--data comparison is summarized graphically in fig.~\ref{fig:xsection}.

It is evident from fig.~\ref{fig:xsection} that $\PW+\Pc$ production at the LHC is a process with great potential for precision physics due to the small uncertainties of both data and NNLO theory.
In particular, the ratio $R_{\PW^\pm\Pj_{\rm c}}$'s PDF uncertainty is much larger than its scale one, which indicates this observable is suitable for constraining PDF sets. The $\Ps-\bar \Ps$ asymmetry is of particular interest.
While its non-vanishing is a clear prediction of QCD \cite{Catani:2004nc} and can be connected to other experimentally verified asymmetries \cite{Czakon:2014xsa}, the absolute size of the $\Ps-\bar \Ps$ asymmetry is at present unclear and appears to be below the current PDF uncertainties.
Various PDF fits take different viewpoints on it.
For example, among the PDF sets we utilize in this work, NNPDF3.1 has non-zero strange asymmetry which results from the independent fitting of $\Ps$ and $\bar \Ps$ while the CT18 family of PDF sets \cite{Hou:2019efy} assumes $\Ps=\bar \Ps$ at a low scale equal to the charm quark pole mass. 

Several factors reduce the sensitivity of $\PW+\Pc$ production to the strange quark asymmetry of the proton.
The contribution mediated by the off-diagonal CKM element $V_{\Pc\Pd}$ is numerically suppressed however it couples at LO in QCD to the $\Pd$ or $\bar \Pd$ quark. The large size of the $\Pd$ valence-quark asymmetry, substantially modifies the ratio $R_{\PW^\pm\Pj_{\rm c}} $.
This is illustrated in Table~\ref{tab:ckm} where the LO predictions for three different PDF sets are presented with the off-diagonal element $V_{\Pc\Pd}$ switched on or off.

Comparing first the predictions computed with NLO PDF sets for $V_{\Pc\Pd}=0$ we observe that the CT18 set leads to $R_{\PW^\pm\Pj_{\rm c}}=1$, as expected. That $R_{\PW^\pm\Pj_{\rm c}}\neq1$ for NNPDF3.1 can be attributed to the fact that $\Ps\neq\bar \Ps$ for this set.
The difference between the two predictions is about $5\%$ which is consistent with the PDF uncertainty at NNLO but is much larger than the scale variation at NNLO.
Once the off-diagonal CKM element is included, the ratio $R_{\PW^\pm\Pj_{\rm c}}$ decreases for both PDF sets. The shift is large, about $7\%$ for CT18 and about $5\%$ for NNPDF3.1, and is comparable to the effect due to $\Ps-\bar \Ps$ asymmetry.
Although the inclusion of NLO and NNLO QCD corrections to the CKM diagonal element $V_{\Pc\Ps}$ reduces the importance of the CKM off-diagonal contribution (see below), a reliable access to the strange asymmetry may require the inclusion of at least the NLO corrections mediated by the off-diagonal CKM element $V_{\Pc\Pd}$.

\begin{table}
  \begin{center}
  \begin{tabular}{cc|c|c|c}
    \toprule
    PDF set & $V_{\Pc\Pd}$ & $\sigma_{\PW^+\Pj_{\rm c}}$ [$\pb$] & $\sigma_{\PW^-\Pj_{\rm c}}$ [$\pb$] & $R_{\PW^\pm\Pj_{\rm c}}$ \\
    \midrule
    \multirow{2}{*}{NNPDF31 LO}  & $=0$     & $9.8395(4)$  & $10.4654(4)$ & $0.94020(5)$ \\
                                 & $\neq 0$ & $12.0725(4)$ & $14.2624(5)$ & $0.84646(4)$ \\
    \midrule
    \multirow{2}{*}{NNPDF31 NLO} & $=0$     & $22.593(2)$  & $23.718(2)$  & $0.95260(6)$ \\
                                 & $\neq 0$ & $24.500(9)$  & $27.29(1)$   & $0.8977(5)$ \\
    \midrule
    \multirow{2}{*}{CT18 NLO}    & $=0$     & $21.675(2)$  & $21.675(2)$  & $1.0000(1)$ \\
                                 & $\neq 0$ & $23.477(9)$  & $25.252(8)$  & $0.9297(5)$ \\
    \bottomrule
    \end{tabular}
  \end{center}
  \caption{\label{tab:ckm}
    LO fiducial cross sections for $\Pp\Pp \to \PW^+\Pj_{\rm c}$, $\Pp\Pp \to \PW^-\Pj_{\rm c}$, and their ratios at the LHC at $\sqrt{s}=7\TeV$ for different PDF sets.
    Predictions for both $V_{\Pc\Pd} = 0$ and $V_{\Pc\Pd} \neq 0$ are given.
    The digit in parenthesis indicates the Monte Carlo statistical error.}
\end{table}

It is clear from the theory--data comparisons in fig.~\ref{fig:xsection} that while not incompatible, the NNLO QCD predictions are in certain tension with data, especially for the plus signature. This difference can be attributed to several factors:
\begin{itemize}
\item In the present computation the c-jets have been defined through a flavored $k_\rT$ algorithm.
This is in contrast with what is done experimentally where anti-$k_\rT$ jets are first reconstructed and then their flavors are identified.
In ref.~\cite{Gauld:2020deh} where $\PZ+\Pb$ has been computed up to NNLO QCD, such effect has been found to be as large as $12\%$.
Nonetheless, it is hard to transfer this number to $\PW+\Pc$ as the processes vary a lot and the measurements are done in different phase spaces using different experimental techniques.
In any case, there is a potential mismatch between data and theory which should be addressed in concert with experimental collaborations in order to perform sound comparisons in the future.

\item As explained above, our best predictions include the effects of $V_{\Pc\Pd} \neq 0$ only at LO in QCD.
Higher-order corrections to this CKM element might slightly modify the present picture. To estimate the effect of higher-order corrections to the off-diagonal CKM matrix element's channel, in Table~\ref{tab:wc} we present the predictions for $\PW+\Pc$ production with $V_{\Pc\Pd}=0$.
By comparing it with Table~\ref{tab:wcCKM} we see that the LO corrections to $V_{\Pc\Pd}$ modify, 
at NNLO, $\sigma_{W^+}$ and $R_{\PW^\pm\Pj_{\rm c}}$ by about $5\%$ and $\sigma_{W^-}$ by about $10\%$. Assuming the pattern of higher order corrections observed for the complete calculation applies to the CKM off-diagonal process, 
we anticipate that the missing higher-order corrections to this subprocess will affect the observable at the level of few per cent. 

\item The inclusion of EW corrections of order $\mathcal{O}\left(\alphas\alpha^3\right)$ and $\mathcal{O}\left(\alpha^4\right)$.
While the former are usually at the level of few per cent for total cross sections, they can become negatively large in the high-energy limit thanks to the effect of Sudakov logarithms.
In ref.~\cite{Denner:2009gj}, the EW corrections have been found to be around $-3\%$ at the level of the fiducial cross section.
Also, the sub-leading corrections of order $\mathcal{O}\left(\alpha^4\right)$ have been found to be slightly below a per cent at the level of the cross section for $\Pp\Pp\to\PZ \Pj$ \cite{Denner:2019zfp}.
All in all, the impact of all types of electroweak corrections should lower the theoretical predictions by few per cent.
Note that we expect the ratio not to be impacted by the EW corrections as the leading Sudakov logarithms factorise and depend only on the external states \cite{Denner:2000jv}.
They should therefore not significantly vary between the two signatures as observed, for example, in ref.~\cite{Chiesa:2019ulk} for same-sign W scattering.

\item Future PDF fits that use this data and NNLO theory might also somewhat modify the above picture.
\end{itemize}

\begin{table}
  \begin{center}
    \begin{tabular}{c|c|c|c}
    \multicolumn{4}{c}{\raisebox{0.2cm}{$V_{\Pc\Pd}=0$}} \\
    \toprule
    Order & $\sigma_{\PW^+\Pj_{\rm c}}$ [$\pb$] & $\sigma_{\PW^-\Pj_{\rm c}}$ [$\pb$] & $R_{\PW^\pm\Pj_{\rm c}} = \sigma_{\PW^+\Pj_{\rm c}}/\sigma_{\PW^-\Pj_{\rm c}}$\\
    \midrule
    LO    & $9.8395(4)^{+11.8\%}_{-15.6\%}$      & $10.4654(4)^{+11.7\%}_{-14.3\%}$ & $0.94020(5)^{+0.93\%}_{-1.46\%}$ \\
    \midrule
    NLO   & $33.266(9)^{+7.8\%}_{-6.9\%}$        & $33.523(9)^{+7.0\%}_{-6.4\%}$   & $0.9923(4)^{+0.74\%}_{-0.54\%}$ \\
    \midrule
    NNLO  & $36.7(1)^{+1.7\% \; +4.0\% ({\rm PDF})}_{-2.9\% \; -4.0\% ({\rm PDF})}$ & $35.7(1)^{+0.7\% \; +4.4\% ({\rm PDF})}_{-2.2\% \; -4.4\% ({\rm PDF})}$ 
    & $1.030(6)^{+0.98\%\; +3.0\% ({\rm PDF})}_{-0.74\%\; -3.0\% ({\rm PDF})}$ \\
    \bottomrule
    \end{tabular}
  \end{center}
  \caption{\label{tab:wc}
    As in Table~\ref{tab:wcCKM} but for $V_{\Pc\Pd}=0$.}
\end{table}

In the rest of this section we consider the behavior of the $\PW+\Pc$ fiducial cross section at different perturbative orders.
In order to quantify the effects of higher-order QCD corrections we consider only the processes mediated by the diagonal CKM matrix element $V_{\Pc\Ps}$. 

We start by comparing the results in Table~\ref{tab:wc} and Table~\ref{tab:wcCKM}. As we already remarked above, the NLO corrections are very large if PDFs of matching order are used. Using a fixed PDF of NNLO accuracy, we observe that the NLO corrections reduce to about $+48\%$ for the plus signature and about $+41\%$ for the minus one.
The NNLO corrections are significantly smaller than the NLO ones and are about $+9\%$ and $+6\%$, for the respective signatures.
The scale variation at both NLO and NNLO is slightly smaller than the scale uncertainty for the case $V_{\Pc\Pd}\neq0$, see Table~\ref{tab:wcCKM}.
This can be explained by the fact that the effect of $V_{\Pc\Pd}\neq0$ is only described at LO and does not benefit from the reduction of scale uncertainty when higher-order effects are included.

\begin{table}[t]
\centering
\begin{tabular}{c|cc|cc|cc}
\multicolumn{7}{c}{\raisebox{0.2cm}{$\Pp\Pp \to \PW^+\Pj_{\rm c}$}} \\
\toprule
\multirow{2}{*}{Contribution}  & \multicolumn{2}{c|}{LO}             & \multicolumn{2}{c|}{NLO}            & \multicolumn{2}{c}{NNLO}  \\
                               & $\sigma$ [$\pb$] & $\delta$ [$\%$] & $\sigma$ [$\pb$] & $\delta$ [$\%$] & $\sigma$ [$\pb$] & $\delta$ [$\%$] \\
\midrule
    $\bar\Ps\Pg$     & $9.8395(4)$ & $100$ & $30.418(9)$            & $91.4$          & $32.07(8)$           & $87.3$ \\
    $\Ps\Pg$         & {-}         & {-}   & {-}                    & {-}             & $9(9)\times10^{-4}$  & $7\times10^{-4}$ \\
    $\Ps\bar\Ps$     & {-}         & {-}   & $0.949(7)\times10^{-2}$& $0.03$          & $3(3)\times10^{-3}$  & $7\times10^{-3}$ \\
    $\bar\Ps\bar\Ps$ & {-}         & {-}   & $2.4(6)\times10^{-4}$  & $7\times10^{-4}$& $-7(2)\times10^{-3}$ & $-0.02$ \\
    $\bar\Ps\Pq$     & {-}         & {-}   & $0.426(1)$             & $1.3$           & $0.66(9)$            & $1.7$ \\
    $\Pq\Pq'$        & {-}         & {-}   & $3.155(2)$             & $9.5$           & $4.83(9)$            & $13.3$ \\
    $\Pg\Pq$         & {-}         & {-}   & {-}                    & {-}             & $0.58(5)$            & $1.6$ \\
    $\Pg\Pg$         & {-}         & {-}   & $-0.741(3)$            & $-2.2$          & $-1.41(8)$           & $-3.8$ \\
\bottomrule
\end{tabular}
\caption{Cross-section contributions according to the initial state at LO, NLO, and NNLO QCD for $\Pp\Pp \to \PW^+\Pj_{\rm c}$ at the LHC at $\sqrt{s}=7\TeV$.
Both the absolute (in $\pb$) and the relative contributions (in per cent) are indicated.
Any quark or anti-quark that is not a strange or anti-strange quark is denoted by $\Pq$ or $\Pq'$.
The digit in parenthesis indicates the Monte Carlo statistical error.
The theoretical predictions are obtained for $V_{\Pc\Pd}=0$.}
\label{tab:contributionsP}
\end{table}
\begin{table}[t]
\centering
\begin{tabular}{c|cc|cc|cc}
\multicolumn{7}{c}{\raisebox{0.2cm}{$\Pp\Pp \to \PW^-\Pj_{\rm c}$}} \\
\toprule
\multirow{2}{*}{Contribution} & \multicolumn{2}{c|}{LO}             & \multicolumn{2}{c|}{NLO}            & \multicolumn{2}{c}{NNLO}  \\
                              & $\sigma$ [$\pb$] & $\delta$ [$\%$] & $\sigma$ [$\pb$] & $\delta$ [$\%$] & $\sigma$ [$\pb$] & $\delta$ [$\%$] \\
\midrule
    $\Ps\Pg$      & $10.4654(5)$& $100$ & $31.996(8)$           & $95.4$  & $33.26(9)$          & $93.2$ \\
    $\bar\Ps\Pg$  & {-}         & {-}   & {-}                   & {-}     & $9(9)\times10^{-4}$ & $-3\times10^{-3}$ \\
    $\Ps\bar\Ps$  & {-}         & {-}   & $7.1(1)\times10^{-3}$ & $0.02$  & $3(2)\times10^{-3}$ & $-7\times10^{-3}$ \\
    $\Ps\Ps$      & {-}         & {-}   & $7.3(1)\times10^{-3}$ & $0.02$  & $4(3)\times10^{-3}$ & $0.01$ \\
    $\Ps\Pq$      & {-}         & {-}   & $0.467(3)$            & $1.4$   & $0.7(1)$            & $1.8$ \\
    $\Pq\Pq'$     & {-}         & {-}   & $1.7792(9)$           & $5.3$   & $2.75(1)$           & $7.7$ \\
    $\Pg\Pq$      & {-}         & {-}   & {-}                   & {-}     & $0.29(6)$           & $0.84$ \\
    $\Pg\Pg$      & {-}         & {-}   & $-0.734(4)$           & $-2.2$  & $-1.25(8)$          & $-3.5$ \\
\bottomrule
\end{tabular}
\caption{As in Table~\ref{tab:contributionsP} but for $\Pp\Pp \to \mu^- \nu_{\mu} \Pj_{\rm c}$.}
\label{tab:contributionsM}
\end{table}

To get a better insight into the interplay of partonic fluxes and higher-order corrections, in Tables~\ref{tab:contributionsP} and \ref{tab:contributionsM} we separately present the contributions to the cross section of each signature from a given initial state. For $\Pp\Pp \to \PW^+\Pj_{\rm c}$ (Table~\ref{tab:contributionsP}), the only contributing channel at LO is $\bar\Ps\Pg$ and most other channels open up at NLO QCD.
While the $\bar\Ps\Pg$ channel is still the dominant one with about $91\%$ of the cross section, it gets diluted mainly by $\Pq'\Pq$ channels which make up almost $10\%$ of the NLO-QCD cross section.
The $\Pg\Pg$ channel which is the driving force to many other LHC processes contributes negatively at a level of about $-2\%$.
At NNLO QCD, the Born channel's relative contribution decreases slightly and represents about $87\%$ of the cross section due to the relative increase of $\Pq'\Pq$ contribution ($13\%$) which receives NLO-QCD corrections at this order.
The $\Pg\Pg$ channel now amounts to about $-4\%$ while all the other channels do not exceed $2\%$ of the NNLO QCD cross section.

Table~\ref{tab:contributionsM} displays a rather similar qualitative picture for the process $\Pp\Pp \to \PW^-\Pj_{\rm c}$.
The main difference is in the relative size of the $\Pq'\Pq$ channels which in this case is only about $5\%$ at NLO.
At NNLO the relative size of this contribution slightly increases to about $8\%$.
As a result, the Born channel ($\Ps\Pg$) has a larger weight at both NLO QCD ($95\%$) and NNLO QCD ($93\%$).
The reason for this behavior is that for the plus signature, the $\Pq'\Pq$ category is dominated by $\Pu\Pq$ contributions.
Due to charge conservation, in the minus case, it is dominated by the $\Pd\Pq$ contributions which are smaller than the $\Pu\Pq$ ones for proton-proton collisions.
The same feature can be observed in the $\Pg\Pq$ channels.
Here the ratio of the plus and minus signatures is close to $2$ as one would expect from the difference between the $\Pd$ and $\Pu$ PDF in the proton. 

\subsection{Differential distributions}

In this section we study various differential distributions and their ratios. We split the discussion in two parts.
We start by comparing our best theoretical predictions to the only differential distribution measured by the ATLAS collaboration.
Then, we present predictions for differential distributions and ratios for which there is no LHC data and study their perturbative behavior.
We note that for all distributions shown in this work the last bins do not include overflow events.

\begin{figure}
        \setlength{\parskip}{-10pt}
        \begin{subfigure}{0.49\textwidth}
                \subcaption{}
                 \includegraphics[width=\textwidth,page=1]{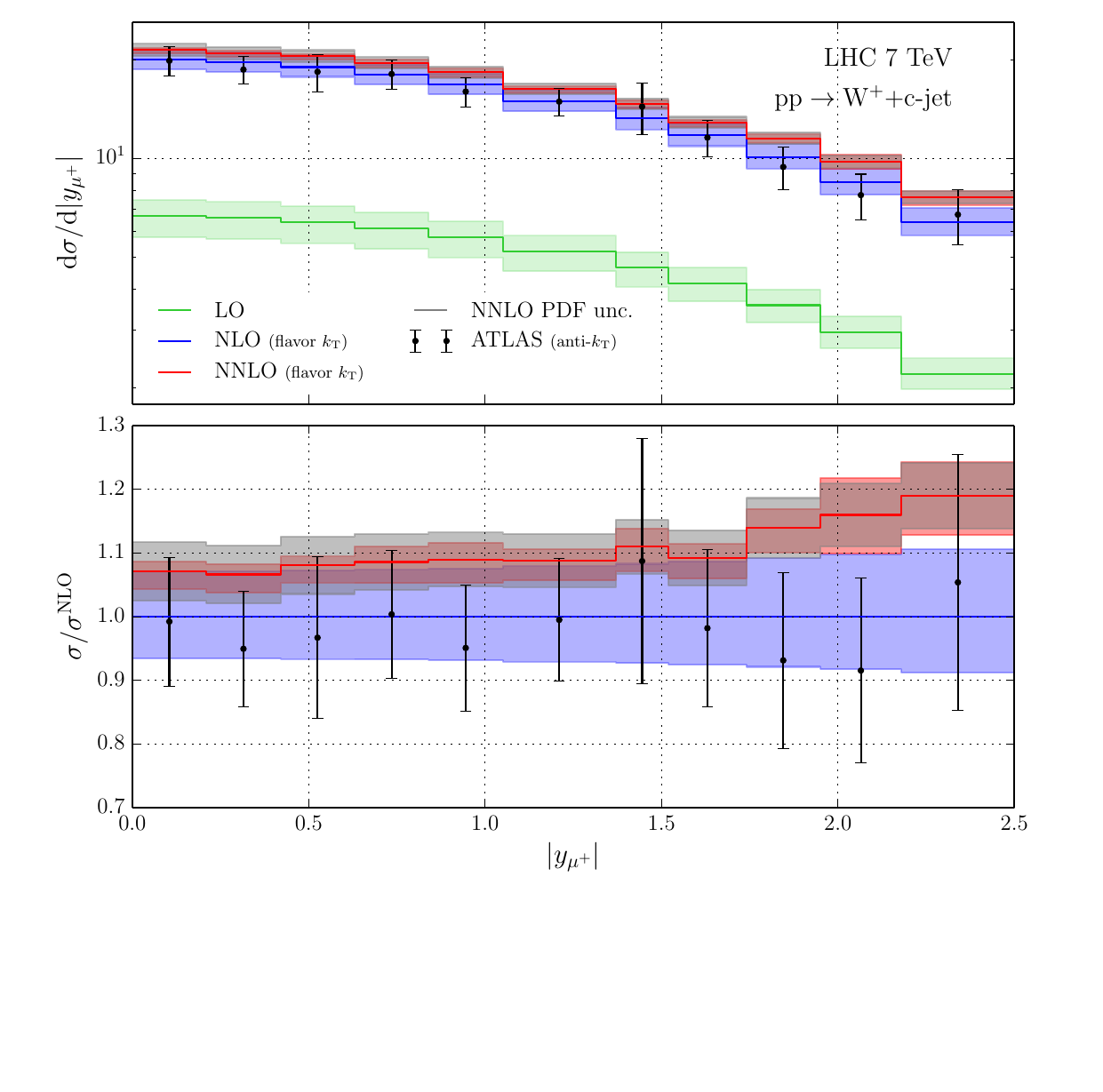}
        \end{subfigure}
        \hfill
        \begin{subfigure}{0.49\textwidth}
                \subcaption{}
                 \includegraphics[width=\textwidth,page=1]{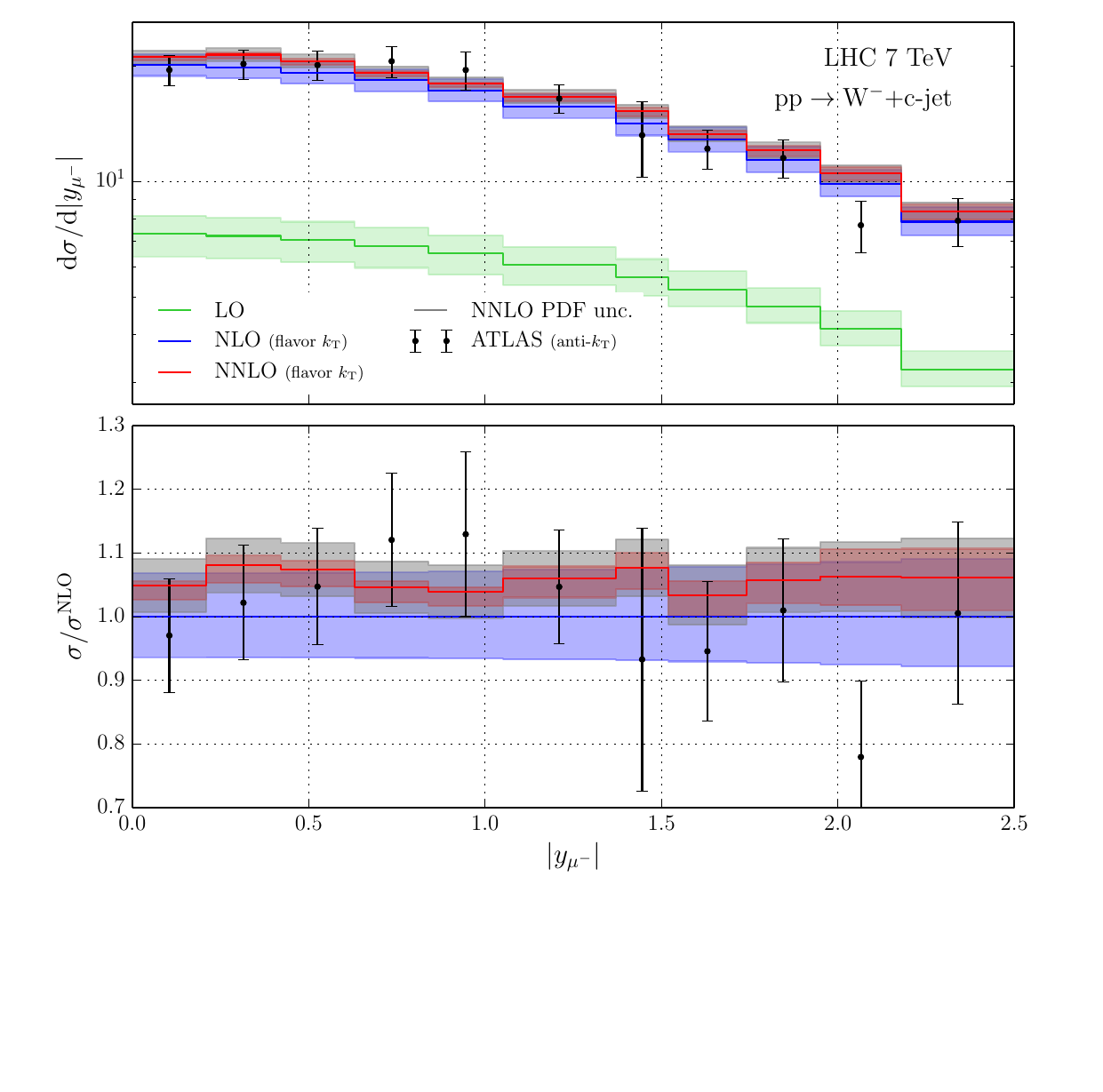}
        \end{subfigure}

        \vspace*{-9ex}
        \caption{\label{fig:data}%
                Differential distributions in the absolute rapidity of the anti-muon in the process 
                $\Pp\Pp \to \PW^+\Pj_{\rm c}$ (left) and of the muon in $\Pp\Pp \to \PW^-\Pj_{\rm c}$ (right) at the LHC with $\sqrt{s}=7\TeV$.
                The upper panel shows the LO, NLO, and NNLO QCD absolute predictions along with ATLAS data \cite{Aad:2014xca}.
                The lower panel displays the same theoretical predictions and data relative to the NLO QCD prediction.
                The grey band represents the PDF variation at NNLO.
                The effect of $V_{\Pc\Pd}\neq0$ is included at Born level only.
                The theory predictions are obtained using the flavor-$k_{\rm T}$ algorithm and
                the ATLAS data has been measured using anti-$k_{\rm T}$.
                }
\end{figure}

In fig.~\ref{fig:data} we show the only observable measured in ref.~\cite{Aad:2014xca}
\footnote{The data of ref.~\cite{Aad:2014xca} have been obtained from {\sc HEPData} \cite{Aad:2014xcaHEPDATA}.}:
the absolute rapidity of the charged lepton for the two signatures. The data is compared to our best theoretical prediction, computed at NNLO in QCD and including the non-diagonal CKM element $V_{\Pc\Pd}$ at LO in QCD.
The agreement is in general relatively good and largely reflects the differences observed at the level of the fiducial cross section.
In particular, no real shape differences are observed as the theoretical predictions at NLO and NNLO QCD are in statistical agreement with the experimental data in most bins.
Nonetheless, despite not being statistically significant, the data seem to be systematically lower than the NNLO QCD prediction for the plus signature while being in agreement for the minus signature.

Possible reasons behind this difference between NNLO theory and data have already been discussed at the cross-section level and they also apply at the differential level. These are 
the difference in the jet algorithms, the lack of higher-order QCD corrections to the off-diagonal CKM matrix element, the absence of EW corrections, and PDF uncertainty.
The role of PDFs appears to be particularly relevant at the differential level since, as can be seen in fig.~\ref{fig:data}, the PDF uncertainty in almost all bins is (significantly) larger than the scale uncertainty.
It is not inconceivable that once the above data has been included in a new global PDF fit the NNLO theory -- data agreement may improve.
The issue of comparing theory and data obtained with different jet algorithms is also rather pressing and, once settled, has the potential for altering the above comparison in a significant way. 

Estimating the effect of higher-order QCD corrections to the non-diagonal CKM elements is slightly more involved here, mostly due to the more pronounced Monte Carlo errors in differential distributions.
At any rate we do not expect them to be very large, in fact, probably within few per cent, given that using same PDF the NLO corrections are $(40-50)\%$.
Finally, in ref.~\cite{Denner:2009gj}, it has been found that for these distributions EW corrections are at the level of $-(2-3)\%$ and no significant shape distortions have been observed across phase space.
Although not a large effect, including such corrections would therefore slightly improve the agreement between theory and data.

Perturbatively, the theoretical prediction for these distributions is fairly well behaved.
It features a reasonably large NNLO/NLO K-factor which is always below $10\%$ for the minus signature while for the plus one it tends to be around $10\%$ in most bins and slowly increases to about $20\%$ for large muon rapidities.
In almost all bins the NLO and NNLO uncertainty bands are compatible.
The difference between the shapes and magnitudes of the K-factors for the two signatures is driven by the relatively significant differences between the partonic fluxes in the two processes, see the discussion towards the end of sec.~\ref{sec:cross-section}.

\begin{figure}[t]
        \setlength{\parskip}{-10pt}
        \begin{subfigure}{0.49\textwidth}
                \subcaption{}
                 \includegraphics[width=\textwidth,page=10]{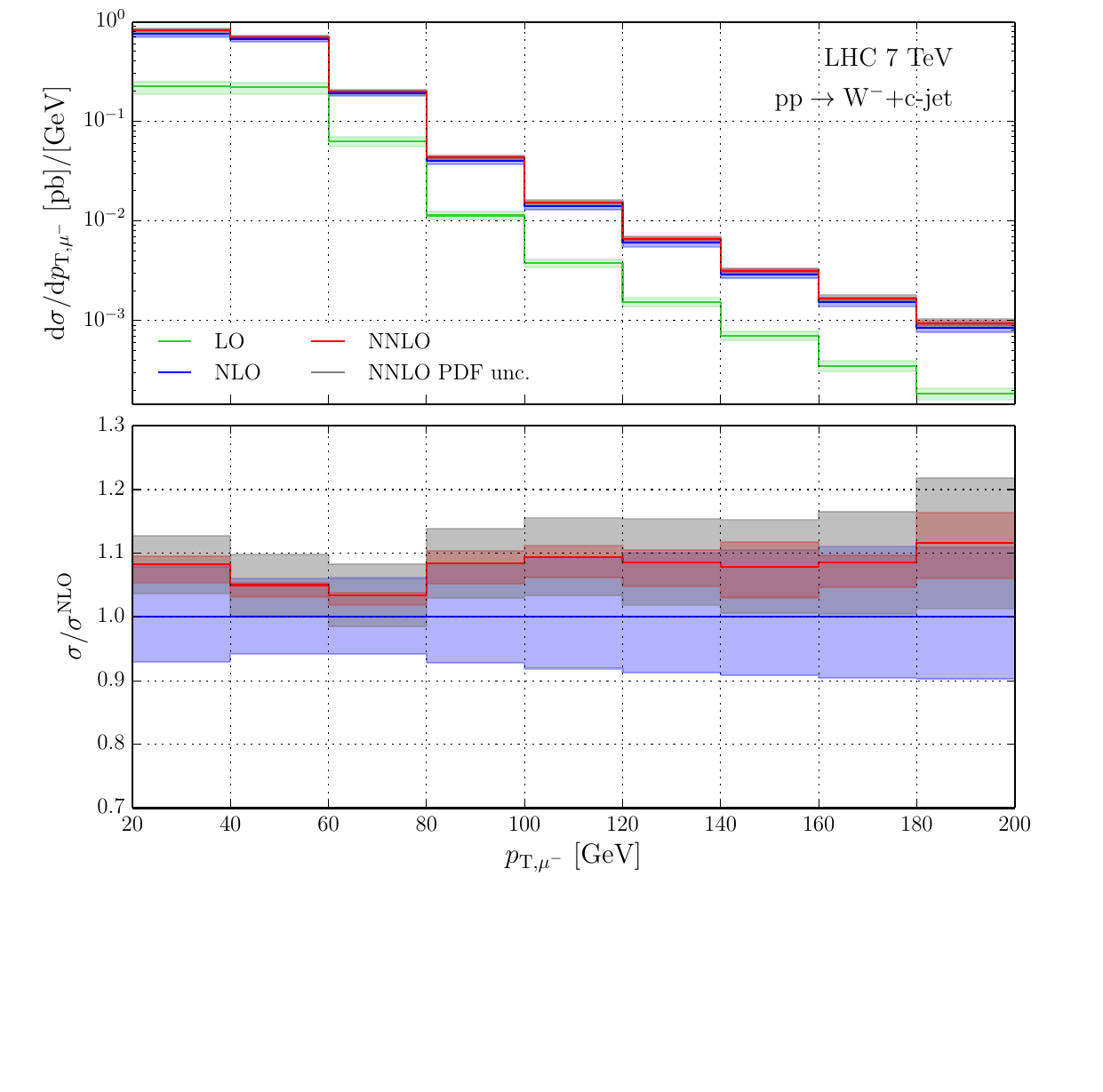}
        \end{subfigure}
        \hfill
        \begin{subfigure}{0.49\textwidth}
                \subcaption{}
                 \includegraphics[width=\textwidth,page=3]{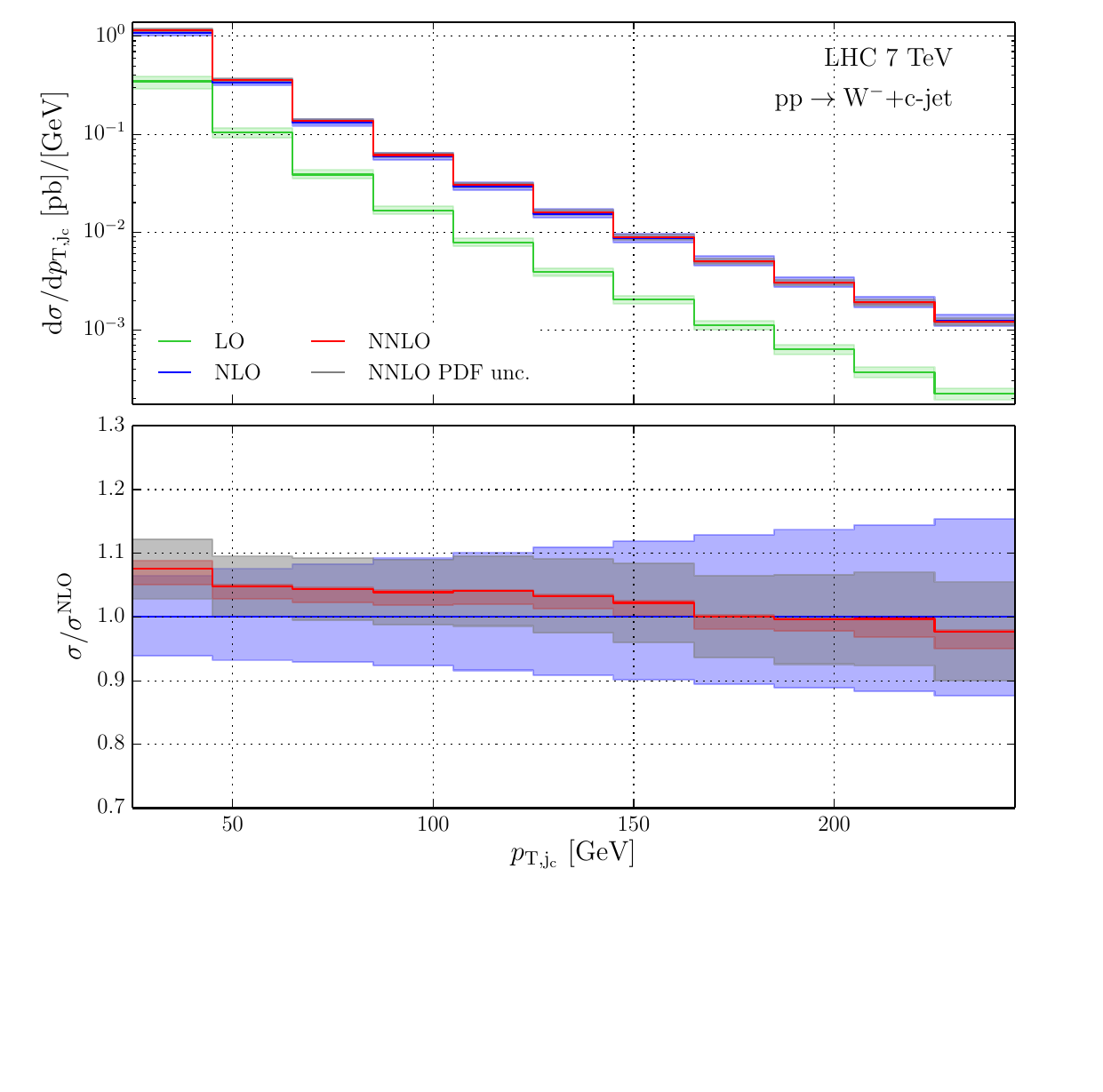}
        \end{subfigure}
        
         \vspace{-4ex}
         
        \begin{subfigure}{0.49\textwidth}
                \subcaption{}
                 \includegraphics[width=\textwidth,page=4]{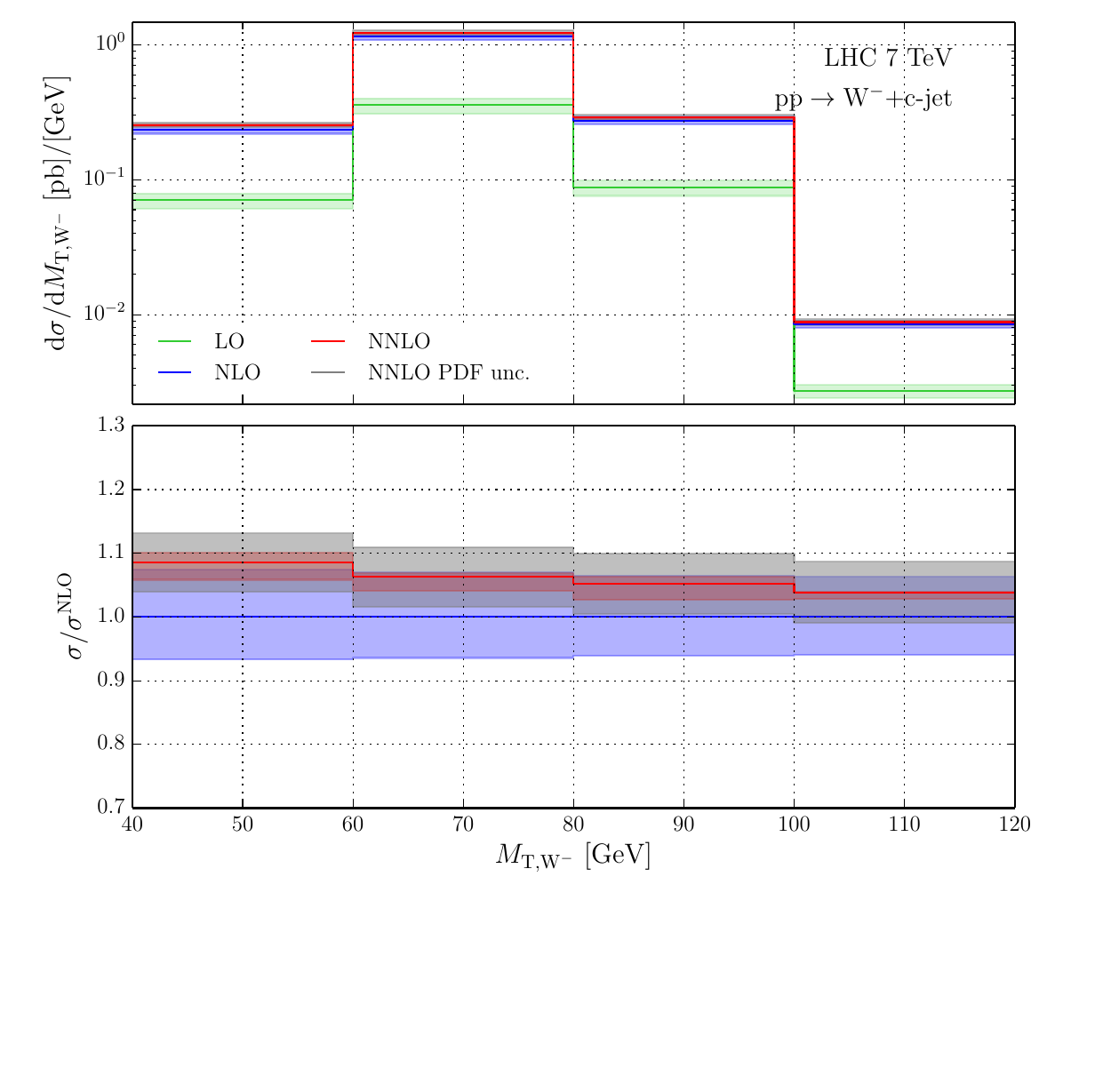}
        \end{subfigure}
        \hfill
        \begin{subfigure}{0.49\textwidth}
                \subcaption{}
                 \includegraphics[width=\textwidth,page=8]{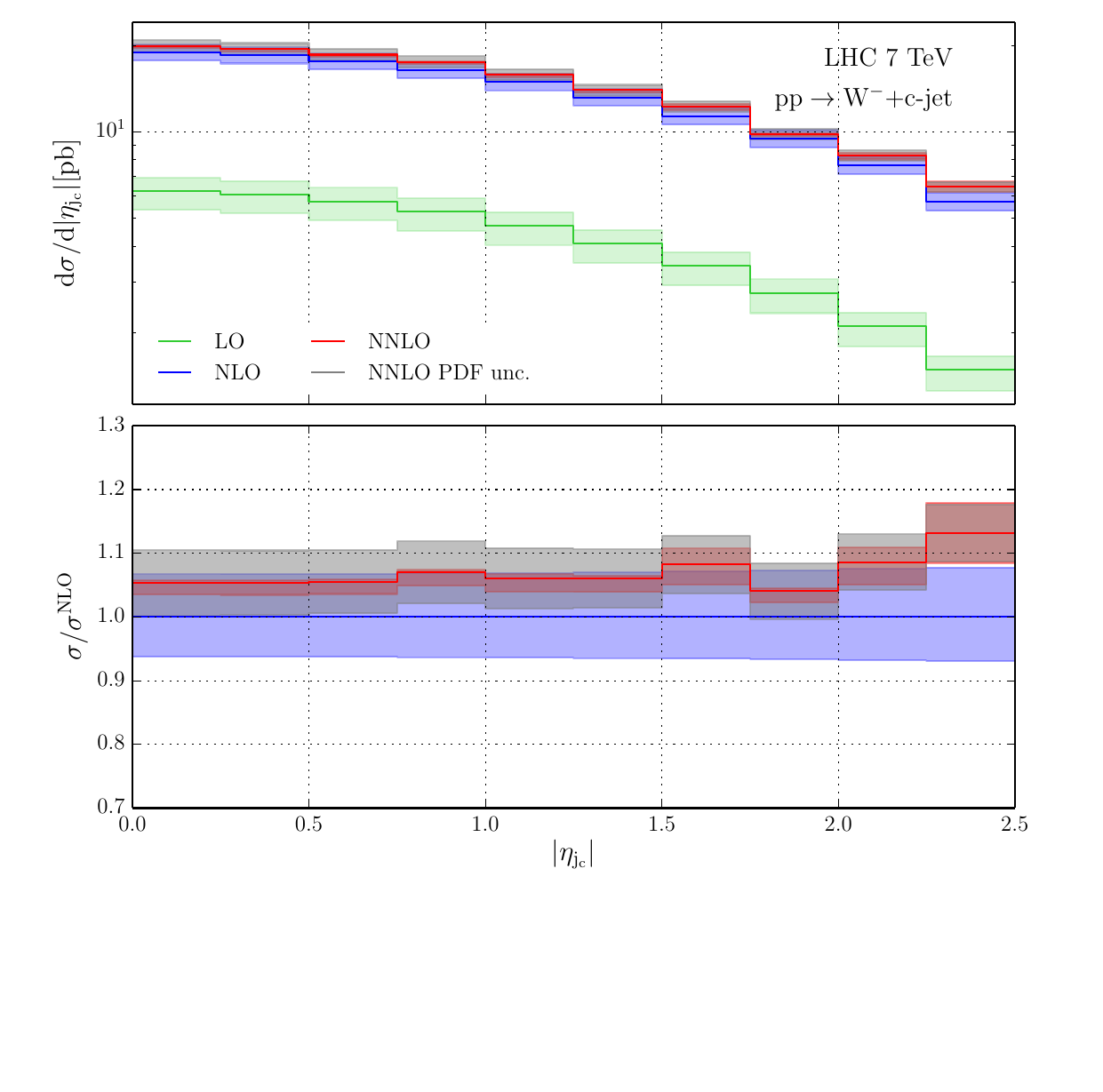}
        \end{subfigure}
        \vspace*{-9ex}
        \caption{\label{fig:pTm}
                Several differential distributions for $\Pp\Pp \to \PW^-\Pj_{\rm c}$ at the $7\TeV$ LHC:
                transverse momentum of the muon~(top left), %
                transverse momentum of the c-jet~(top right),
                transverse mass of the W boson~(bottom left), and
                pseudo-rapidity of the c-jet~(bottom right).
                The upper panels show the absolute distributions in LO, NLO, and NNLO QCD.
                The lower panels display the same theoretical predictions but normalised to the central NLO prediction.
                The grey band represents the PDF variation at NNLO.
                The green/blue/red bands represent scale uncertainties.
                The theoretical predictions are obtained for $V_{\Pc\Pd}=0$.}
\end{figure}

\begin{figure}
        \setlength{\parskip}{-10pt}
        \begin{subfigure}{0.49\textwidth}
                \subcaption{}
                 \includegraphics[width=\textwidth,page=7]{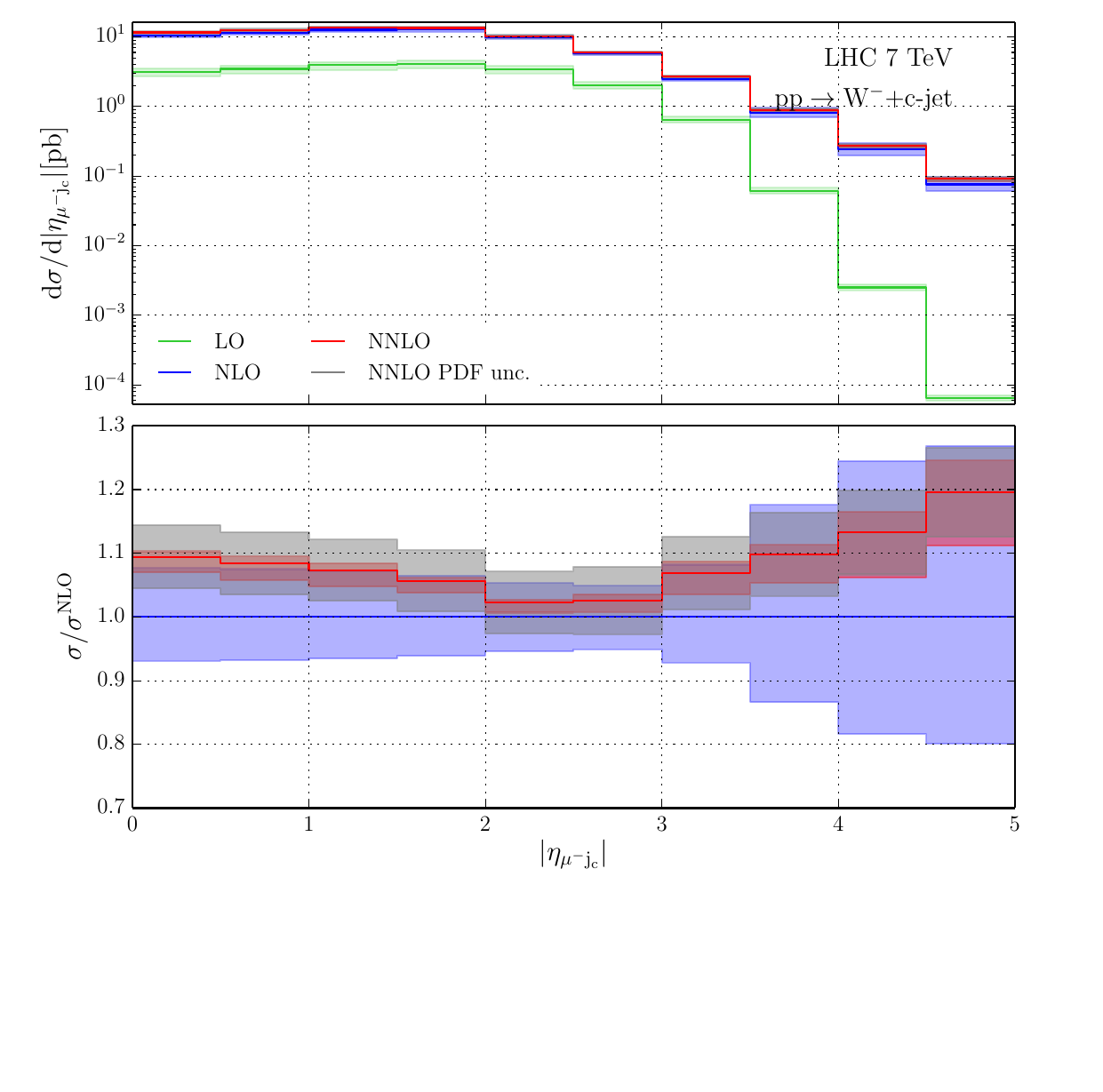}
        \end{subfigure}
        \hfill
        \begin{subfigure}{0.49\textwidth}
                \subcaption{}
                \includegraphics[width=\textwidth,page=9]{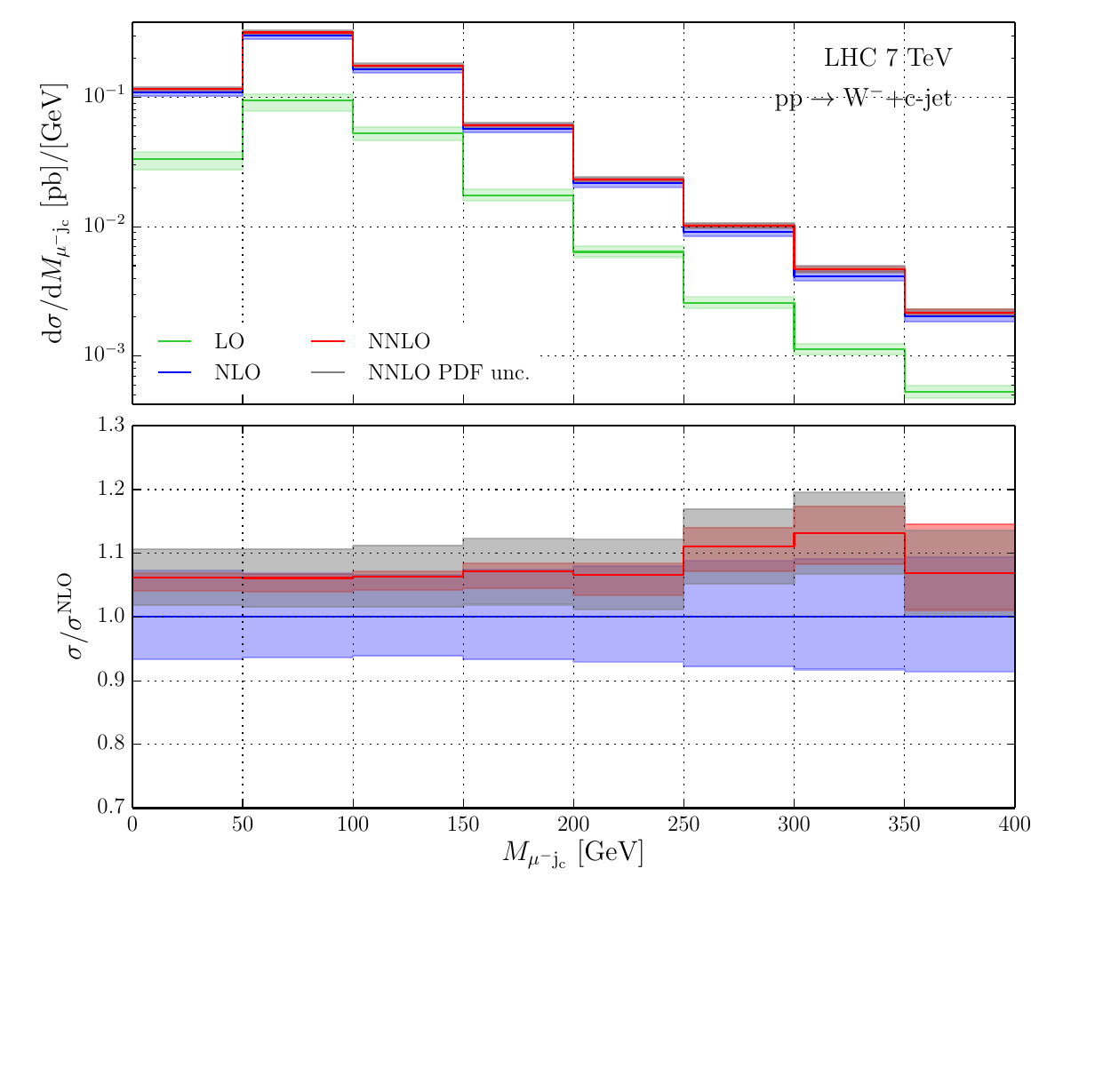}
        \end{subfigure}
        
         \vspace{-4ex}
         
        \begin{subfigure}{0.49\textwidth}
                \subcaption{}
                 \includegraphics[width=\textwidth,page=6]{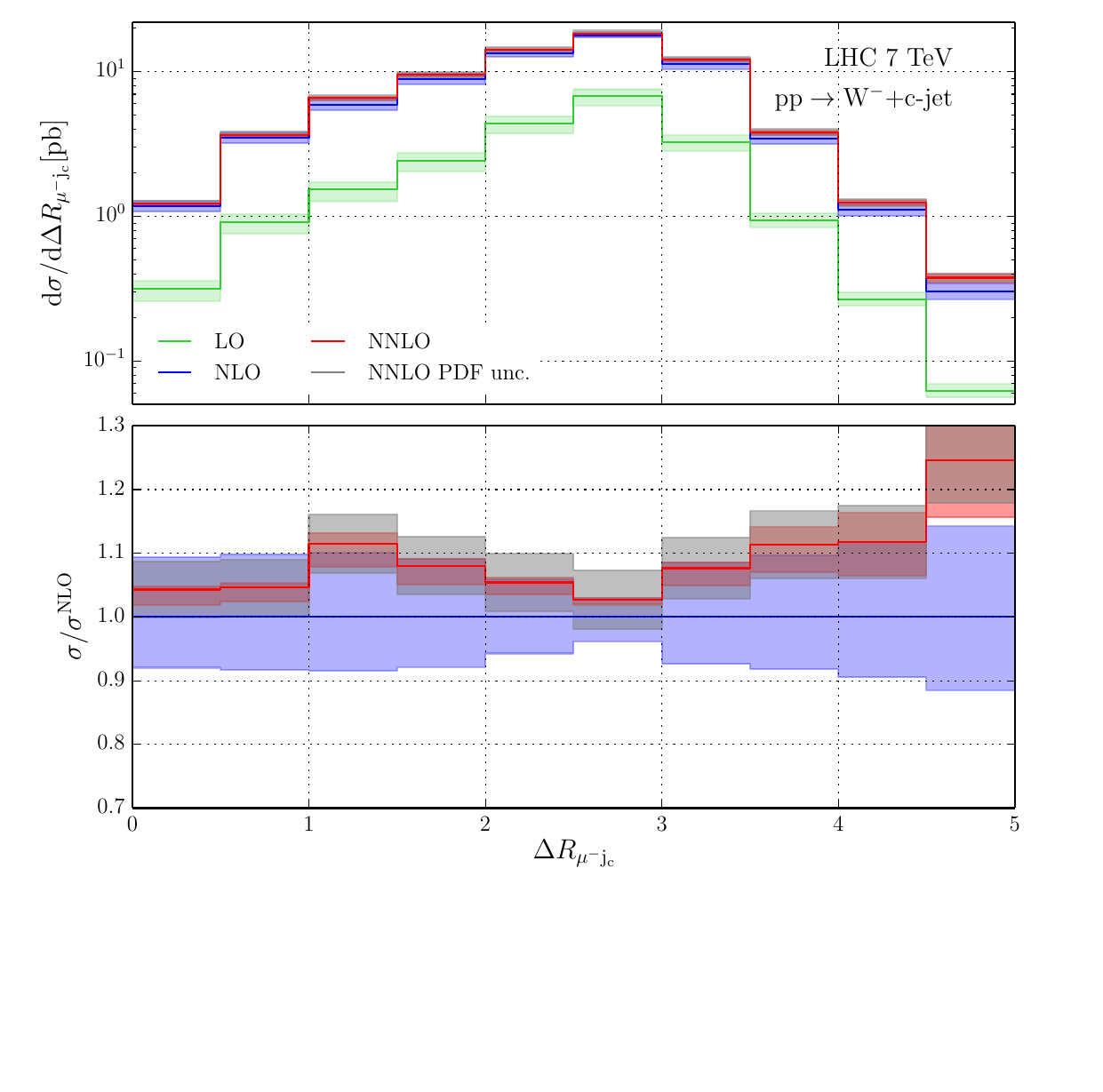}
        \end{subfigure}
        \hfill
        \begin{subfigure}{0.49\textwidth}
                \subcaption{}
                 \includegraphics[width=\textwidth,page=11]{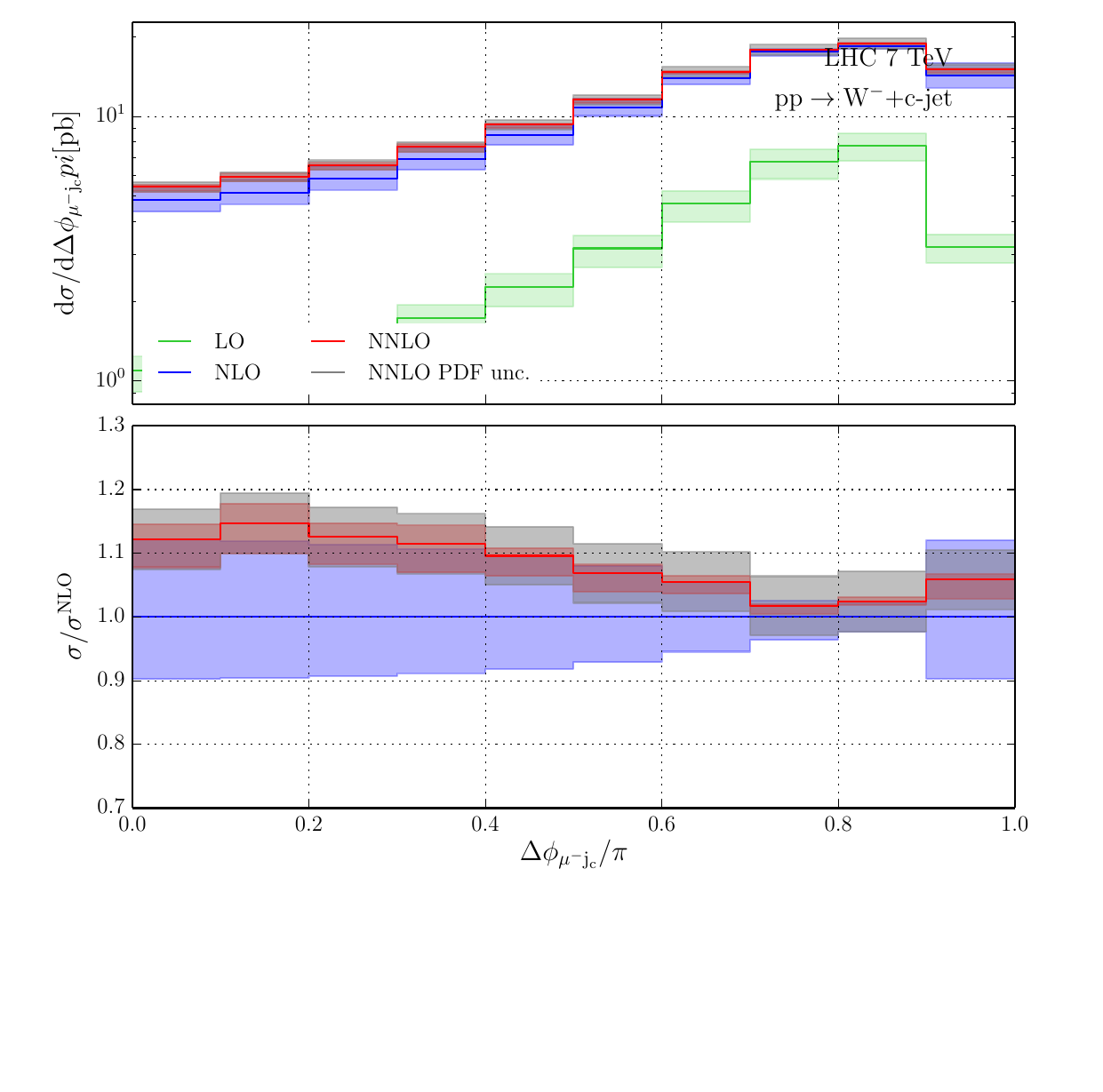}
        \end{subfigure}
        \vspace*{-9ex}
        \caption{\label{fig:rapm}
        As in fig.~\ref{fig:pTm} but for the
                pseudo-rapidity of the muon and c-jet system~(top left),
                invariant mass of the muon and c-jet~(top right),
                rapidity--azimuthal-angle distance between the muon and c-jet~(bottom left), and
                azimuthal angle between the muon and c-jet~(bottom right).}
\end{figure}

\begin{figure}
        \setlength{\parskip}{-10pt}
        \begin{subfigure}{0.49\textwidth}
                \subcaption{}
                 \includegraphics[width=\textwidth,page=10]{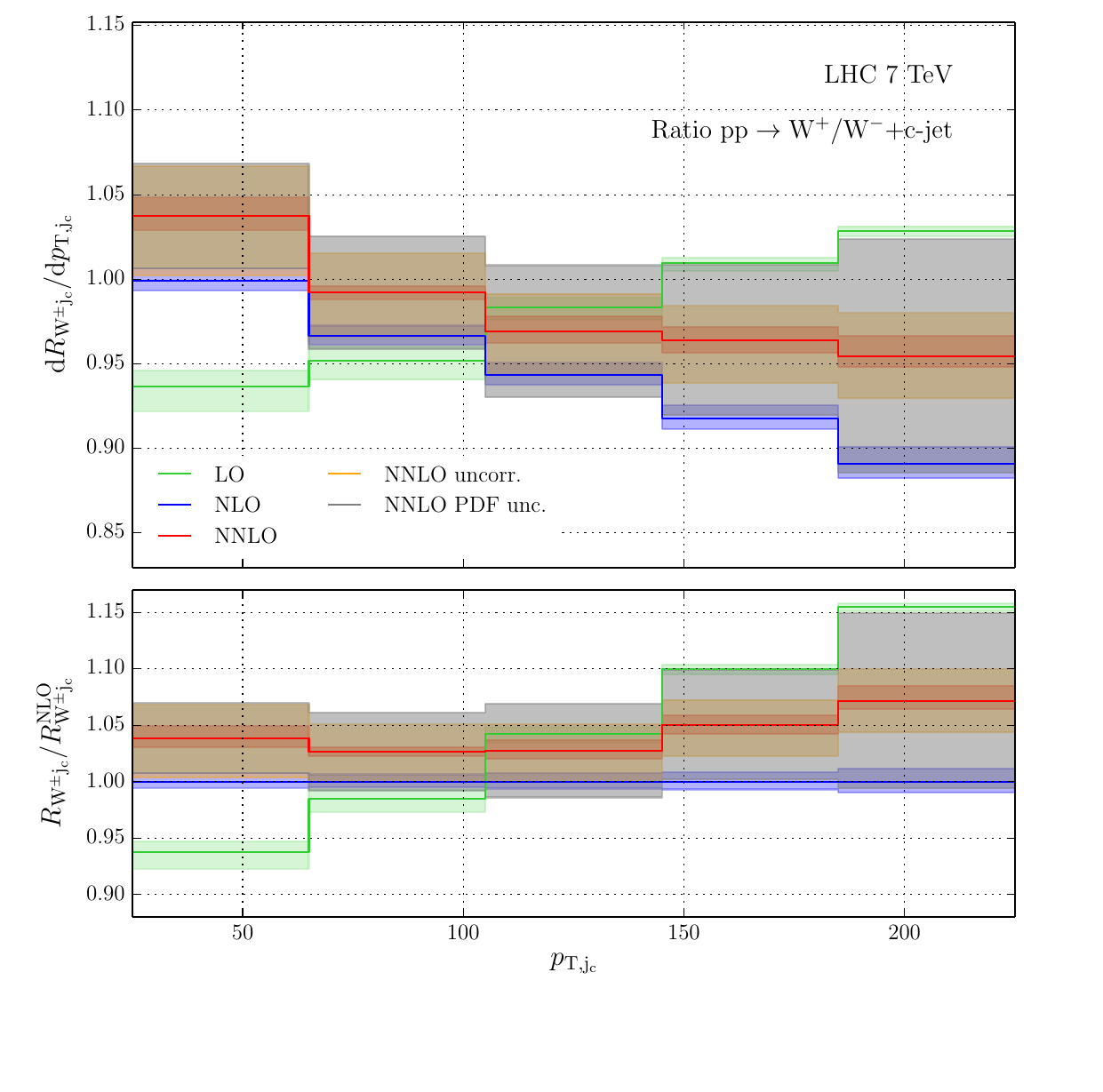}
        \end{subfigure}
        \hfill
        \begin{subfigure}{0.49\textwidth}
                \subcaption{}
                 \includegraphics[width=\textwidth,page=13]{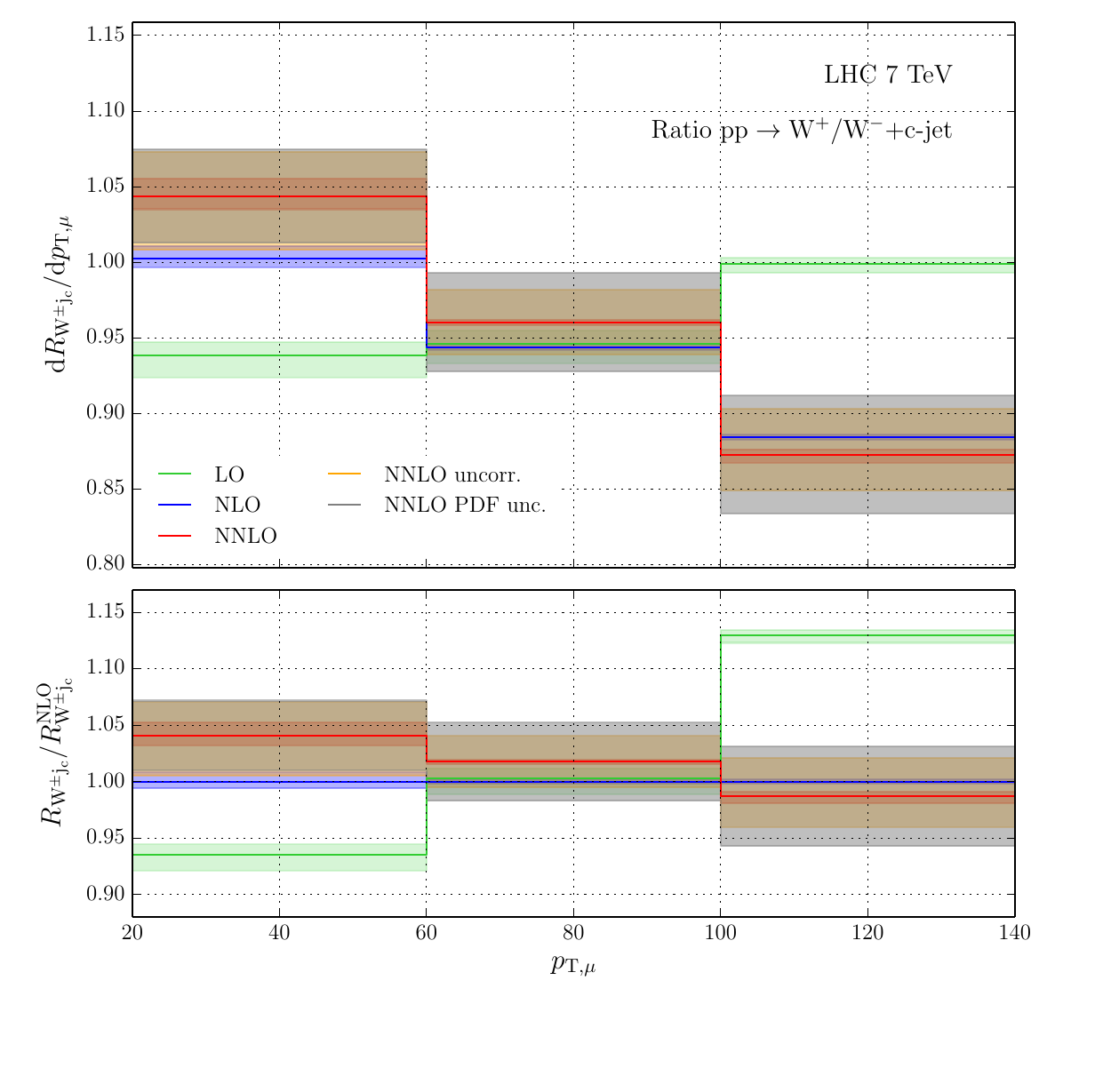}
        \end{subfigure}

        \begin{subfigure}{0.49\textwidth}
                \subcaption{}
                 \includegraphics[width=\textwidth,page=12]{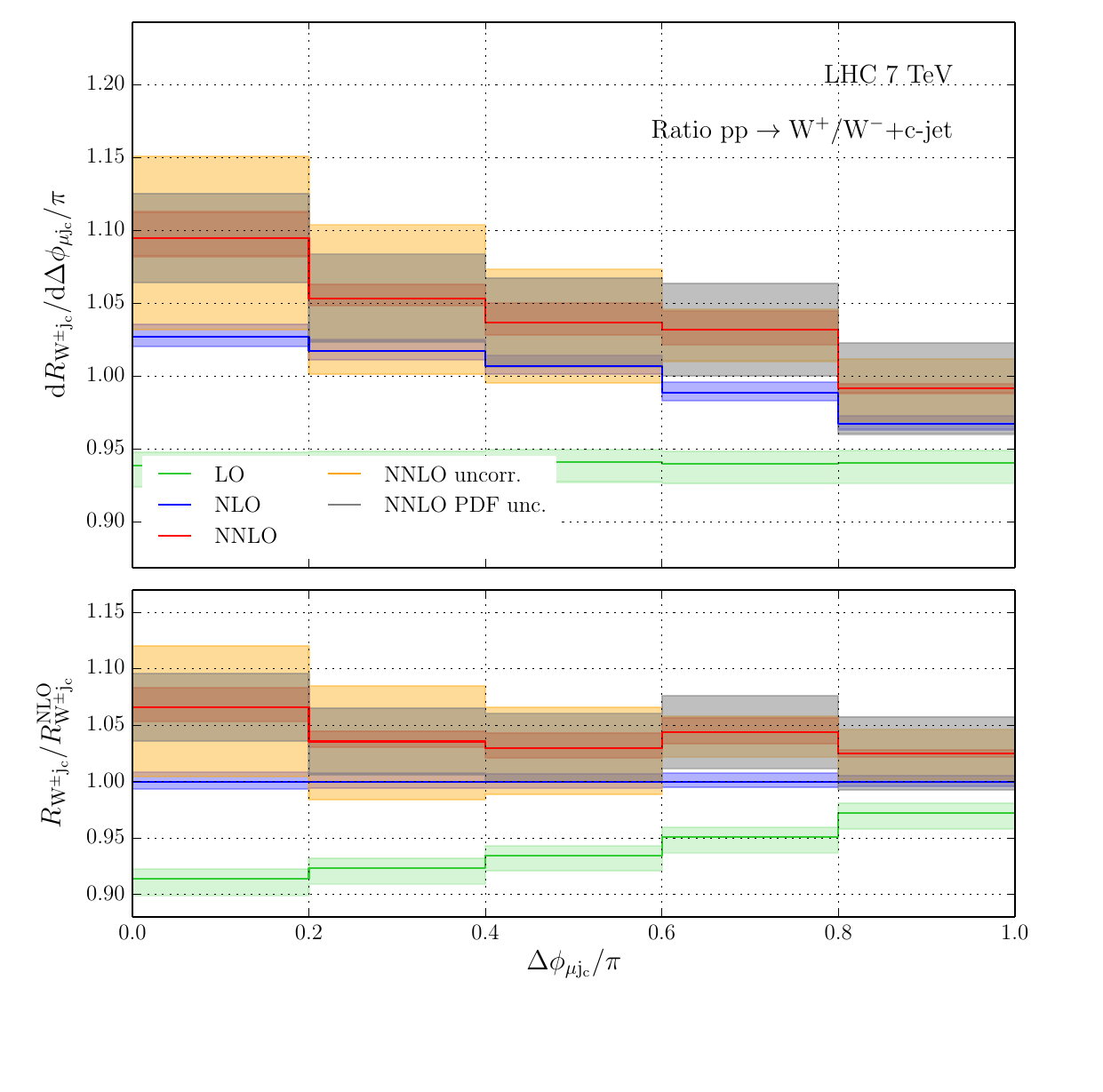}
        \end{subfigure}
        \hfill
        \begin{subfigure}{0.49\textwidth}
                \subcaption{}
                 \includegraphics[width=\textwidth,page=9]{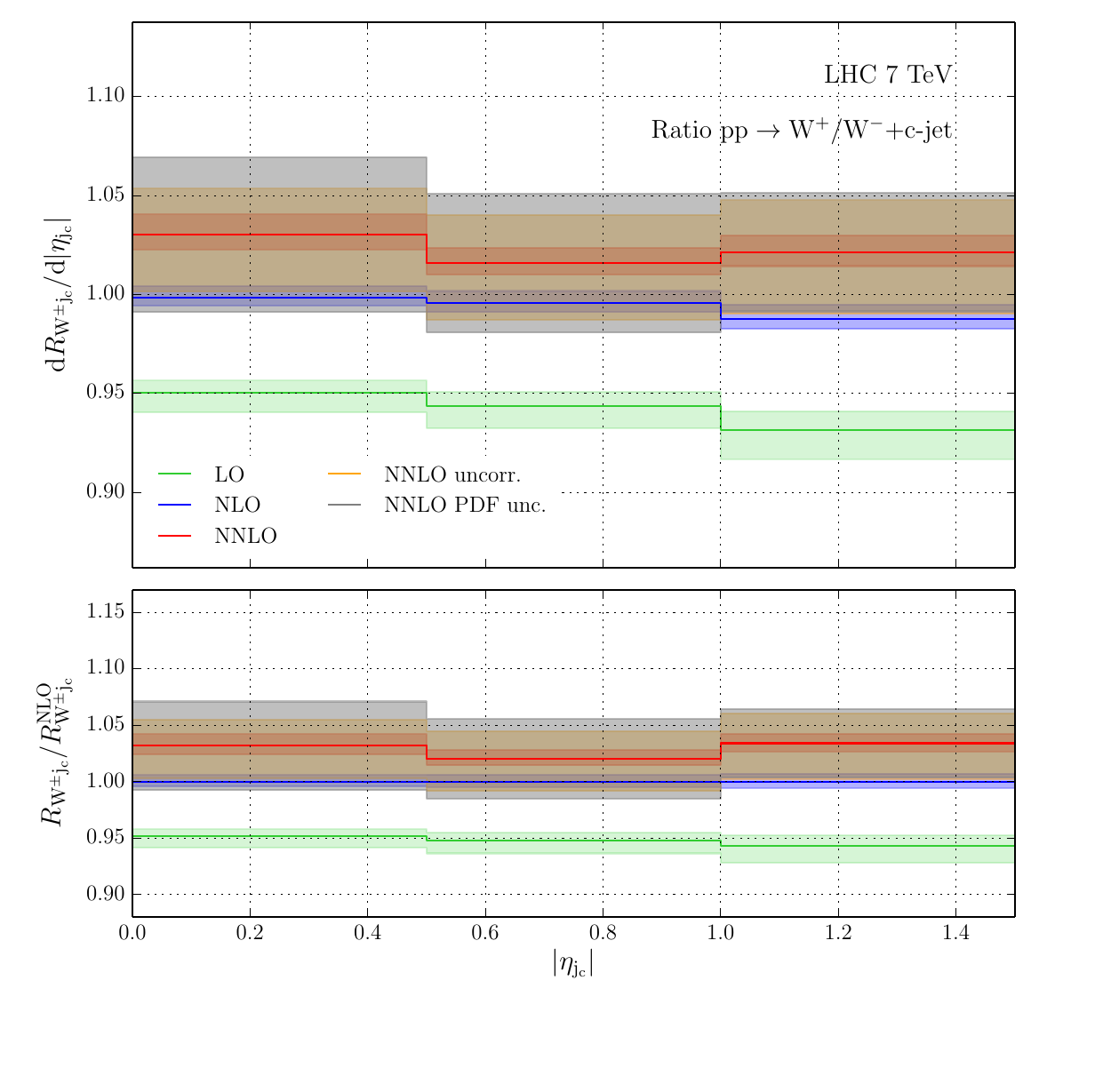}
        \end{subfigure}

        \vspace*{-3ex}
        \caption{\label{fig:ratio}%
                Ratios of differential distributions in $\Pp\Pp \to \PW^+\Pj_{\rm c}$ and $\Pp\Pp \to \PW^-\Pj_{\rm c}$ at the $7\TeV$ LHC:
                transverse momentum of the c-jet~(top left), %
                transverse momentum of the (anti-)muon~(top right),
                azimuthal angle between the (anti-)muon and c-jet~(bottom left), and
                pseudo rapidity of the c-jet~(bottom right).
                The upper panel shows the ratios of the absolute distributions in LO, NLO, and NNLO QCD.
                The lower panel displays the same theoretical predictions but normalised to the central NLO one.
                The grey band represents the PDF variation at NNLO.
                The green/blue/red bands represent correlated scale uncertainties 
                while the orange one is for the uncorrelated 31-point prescription at NNLO.
                The theoretical predictions are obtained for $V_{\Pc\Pd}=0$.}
\end{figure}

In the rest of this section we consider a number of differential distributions for which presently there is no data.
Since our aim in the following is to exhibit the behavior of QCD corrections in $\PW+\Pc$ production at various perturbative orders, all distributions discussed in the rest of this section do not include the contribution proportional to $V_{\Pc\Pd}$.
In Figs.~\ref{fig:pTm} and \ref{fig:rapm} we show differential distributions for the process $\Pp\Pp \to \PW^-\Pj_{\rm c}$ while in fig.~\ref{fig:ratio} we present a selection of differential ratios between the two processes.

In figs.~\ref{fig:pTm} and \ref{fig:rapm} we show the theoretical predictions for eight differential distributions.
For each distribution we show the absolute predictions in LO, NLO and NNLO QCD as well as the NNLO/NLO K-factor.
We do not show the NLO/LO K-factor since, as we already mentioned in our discussion of the fiducial cross section, the LO approximation for this process is rather poor due to the rather small LO PDF. 

The scale variations at NLO and NNLO as well as the sizes of the NNLO/NLO K-factors are rather similar across all differential distributions.
The NLO scale variation is relatively small, well within $10\%$ for most bins and exceeds $10\%$ for only two distributions:
for the transverse momentum of the c-jet the NLO scale variation is about $5\%$ at low $p_\text{T}$ and slowly increases to about $15\%$ at $p_\text{T}$ about $250\GeV$ while for the pseudo-rapidity of the muon and c-jet system it decreases from $\eta=0$ to about $\eta=2.5$ after which is starts to increase rather fast and exceeds 20\% at $\eta=5$. 

The reduction of scale uncertainty when going from NLO to NNLO is substantial.
In most bins it is about a factor of 3 while in many bins it can be as large as 4.
The largest reduction in scale uncertainty is observed in the $p_\text{T}$ distribution of the c-jet which displays scale variation that is as small as couple of percent and is almost independent of $p_\text{T}$.
This is quite remarkable given the PDF variation of the NNLO prediction is several times larger than the scale variation.
This means that the $p_\text{T}$ distribution of the c-jet can be an excellent place for constraining PDF sets with only very small impact from unknown terms beyond NNLO.
In fact, remarkably, one observes that in all distributions the NNLO PDF variation surpasses the scale one - by a factor of two in most bins - and only occasionally the two variations are about equal.

In addition to the very significant decrease in scale variation when going from NLO to NNLO, the inclusion of higher order corrections in the W+c-jet process leads to non-trivial modifications of the shapes of distributions (\emph{i.e.}\ the NNLO/NLO K-factor has non-trivial shape).
For almost all distributions and bins the NNLO scale uncertainty band is within the NLO one, although it tends to be close to NLO band's upper edge. 
This means that the NNLO/NLO K-factor is positive and moderate, typically about 5--10\%.
The only distribution that shows negative NNLO/NLO K-factor is the $p_\text{T}$ of the c-jet which becomes zero at about 180 GeV and steadily decreases towards negative values as $p_\text{T}$ increases. 

Among the differential distributions discussed here, the transverse mass of the W boson, defined in eq.~(\ref{eq:mTW}), is of particular interest since it acts as a proxy for the W invariant mass.
It peaks just below $80\GeV$ and features rather stable corrections at both NLO and NNLO QCD, especially in the range around the peak of the distribution.
The NNLO corrections are largest for low values of the transverse mass of the W boson and slowly decrease as $M_{\rm T,W}$ increases, at least in the kinematic range considered in this work.
The scale variation is very small in the area below $100\GeV$ while the PDF variation there is significantly larger than the scale one.


The four distributions shown in fig.~\ref{fig:rapm} feature the muon and c-jet system which constitutes the observable part of the Born final state.
A notable feature of these distributions is the presence of very large NLO/LO corrections for certain (extreme) kinematics.
For example, the NLO correction of the absolute pseudo-rapidity of the muon and c-jet system beyond $\eta=2.5$ becomes several orders of magnitudes larger than the LO one.
This effect is partly due to the large difference between LO and NLO PDFs but also because at the Born level such kinematics can only be satisfied by very boosted events which are very much suppressed with respect to back-to-back topologies at low pseudo-rapidities (below $\eta=2.5$).
At NLO and higher orders such configurations can easily be obtained through extra real radiations that recoils against the muon and c-jet system. 
Similar feature is also observed in the distribution of the azimuthal angle between the muon and the c-jet.
At LO this distribution reaches its maximum slightly below $\Delta\phi=\pi$ since the c-jet recoils against the W boson and not against the charged lepton.
The energy carried away by the neutrino is thus responsible for this behavior.
When extra real QCD radiation is present, configurations where the c-jet and the muon are perfectly aligned thus become more probable which leads to the large NLO QCD corrections observed in the last bin of that distribution.

In the rest of this section we study the behavior of four differential ratios. Shown in fig.~\ref{fig:ratio} are the absolute ratio distributions together with their K-factors.
The most striking feature in all cases is the reduction of scale uncertainty in the ratio relative to the scale uncertainty of the corresponding distributions.
This reduction of scale uncertainty is so large that for all distributions and bins the three perturbative orders included here are never overlapping and, typically, are relatively far from each other.
While this reduction in scale variation is well known from other ratios studied in collider physics, the lack of overlap between the various orders does raise the question of the reliability of the estimate of ratio's uncertainty due to missing higher order corrections. 

The above is a valid concern; still, from the information available in this work one can indirectly conclude that the NNLO prediction for the ratio may not be far from the \emph{true} theoretical prediction.
The reason for this is the observation that the absolute differential distributions for a given signature show good convergence at NNLO, with the NNLO/NLO K-factor typically being below $10\%$.
Assuming perturbative convergence holds in this process (we see nothing in this process that may suggest otherwise) one can conclude that the corrections beyond NNLO would be at the percent level which, in turn, implies that the N$^3$LO corrections to the ratios would most likely be consistent with the NNLO ones. 

The higher-order corrections to the ratios displayed in fig.~\ref{fig:ratio} have important impact on the normalizations and shapes of ratios.
The fact that the NNLO and NLO predictions are only marginally closer to each other than the NLO and LO ones should likely not be viewed as lack of perturbative convergence.
Indeed, this situation should be contrasted with the corresponding K-factors for the absolute distributions where the NLO/LO K-factor is typically huge. 
In other words the ratio appears to be much more well behaving than the absolute distributions and the only reason this is not apparent in fig.~\ref{fig:ratio} are the extremely small scale variations at all orders.
The sizes of the ratios' K-factors are fairly small; mostly within $5\%$ at both NLO and NNLO and in no case they exceed $10\%$.
This is consistent with the behavior of the ratio of the fiducial cross sections shown in fig.~\ref{fig:xsection}.

Another, even more interesting, feature of the ratios shown in fig.~\ref{fig:ratio} are the sizes of their PDF uncertainties.
As for the absolute distributions discussed previously, we have shown in fig.~\ref{fig:ratio} the PDF uncertainty at NNLO only.
Relative to the correlated scale variation, the PDF one is simply huge and in some bins exceeds the scale variation by up to an order of magnitude.
The inclusion of the PDF variation therefore has an important impact on any conclusion about the reliability of the theory predictions.
Indeed, once PDF uncertainty is included, the NLO and NNLO predictions become consistent in almost all bins within the uncertainty envelope of the NNLO predictions.
This observation further suggests that the various perturbative orders may be made consistent within their scale variations for some new modified PDF set that differs with the current one within the latter's PDF uncertainty.
It is this feature that makes it evident that $\PW+\Pc$ production has significant potential for improving existing PDF sets.
The differential ratio of the two signatures is particularly well suited for this task. 

As for the fiducial cross-section, in fig.~\ref{fig:ratio} we present differential predictions for the cross-sections ratio with scale variation in each bin derived from the uncorrelated 31-point method.
For clarity of presentation we only show the uncorrelated band for the NNLO case.
The size of the uncorrelated scale bands at LO and NLO can easily be estimated from the bands of the corresponding differential cross-sections.
As can be expected the uncorrelated scale variation is significantly larger than the correlated one.
Interestingly, in most bins the uncorrelated scale variation is smaller than, or similar to, the PDF uncertainty.

Lastly, we would like to address a potential subtlety related to the determination of the PDF uncertainty of the ratios shown in fig.~\ref{fig:ratio}. 
As explained earlier in this work, we utilize not the full PDF set but a reduced set of modified PDFs.
Since these sets are different for the two signatures, and we intend to compute {\it correlated} PDF uncertainties, one may worry if the procedure suggested in ref.~\cite{Carrazza:2016htc} (that we follow here) fails, and does not correctly produce the PDF variation of the ratio.
To check that this is not the case we have compared the NLO predictions for the above ratios convoluted, in turn, with the standard and with the reduced sets of NNLO PDFs.
Comparing the two predictions, at the level of the fiducial cross section and of the differential ratio in the c-jet's $p_T$, we observe that the two PDF variations are extremely close to each other and only show noticeable (but immaterial) difference in the last bin of that differential ratio.
From this check we conclude that the size of the PDF uncertainty of the ratios shown in fig.~\ref{fig:ratio} is not artificially inflated in a significant way as a result of the PDF error calculation and, likely, is a real effect.

\section{Conclusions}\label{sec:conclusions}

The production of a W boson in association with a charm jet at the LHC represents a very sensitive probe of the proton's strange quark content.
To enable this PDF's precision extraction, in this article we calculate for the first time the NNLO QCD corrections to this process.
We consider both signatures $\Pp\Pp\to \mu^+\nu_\mu\Pj_{\rm c}$ and $\Pp\Pp\to \mu^-\bar\nu_\mu\Pj_{\rm c}$.
We present comprehensive predictions for fiducial cross sections, differential distributions and (differential) ratios of the two signatures.
In addition to the dominant Born process which is included in NNLO QCD, our best predictions include at LO in QCD the process mediated by the off-diagonal CKM matrix element $V_{\Pc\Pd}$. 

The NNLO QCD corrections are typically around $10\%$ which is larger than the NNLO corrections found in $\PW$ production with an inclusive jet.
The inclusion of the NNLO corrections leads to a significant reduction of the theoretical uncertainty due to missing higher-order terms.
A remarkable feature of this process is that at this order, the PDF uncertainty is consistently larger than the scale one.
For this reason this process offers excellent opportunities for high-quality fitting of the strange quark PDF and possibly even the $\Ps\bar \Ps$ asymmetry of the proton.

Our best theoretical predictions have been compared to existing $7\TeV$ ATLAS measurements of the fiducial cross section and of the rapidity of the charged lepton.
The data for the plus signature tends to be lower than the NNLO QCD predictions.
We have discussed three sources for this discrepancy:
the different jet algorithms used in our computation and in the experimental analysis, EW corrections which are not accounted for in this work, and the need for inclusion of higher-order QCD corrections to quark-mixing effects mediated by off-diagonal CKM matrix elements.
The difference between the jet algorithms has the potential to be numerically significant and highlights a problem that has been outstanding for a long time:
the effective description of flavored jets and the mutual consistency between NNLO theory and LHC measurements.
We hope our work will serve as an added motivation for finding a suitable resolution to this discrepancy.
Its satisfactory resolution will open the door for high-precision analyses of flavored jet production at the LHC and for the NNLO QCD extraction of strange quark PDFs from LHC data.
Finally, we hope our work will motivate extending the capabilities of existing libraries for the numerical evaluation of tree-level and one-loop amplitudes by including also the calculation of quark-mixing effects in LHC processes.

\section*{Acknowledgements}

The authors would like to thank Robert Thorne for suggesting this study, Shayan Iranipour and Maria Ubiali for related discussions and Zahari Kassabov for providing the reduced NNLO PDF sets.
The authors thank the NNLOJet collaboration and in particular Alexander Huss for information regarding the set-up of ref.~\cite{Gehrmann-DeRidder:2019avi} as well as the ATLAS collaboration for clarifications regarding the $\PW+\Pc$ analysis.
M.P. thanks Thomas Gehrmann and Ettore Remiddi for help in using {\sc hplog} and {\sc tdhpl}.
The work of M.C. was supported by the Deutsche Forschungsgemeinschaft under grant 396021762 -- TRR 257.
The research of A.M., M.P., and R.P. has received funding from the European Research Council (ERC) under the European Union's Horizon 2020 Research and Innovation Programme (grant agreement no.~683211). A.M. was also supported by the UK STFC grants ST/L002760/1 and ST/K004883/1.
M.P. acknowledges support by the German Research Foundation (DFG) through the Research Training Group RTG2044.

\bibliographystyle{utphys.bst}
\bibliography{wc_nnlo}
\end{document}